\documentclass[12pt,letterpaper,preprint]{aastex}

\slugcomment{Draft Version 12}

\shorttitle{Massive star formation in the Norma spiral arm II}
\shortauthors{Chavarr\'{\i}a et al.}

\newcommand{\be}{\begin{equation}}
\newcommand{\ee}{\end{equation}}
\newcommand{\bd}{\begin{displaymath}}
\newcommand{\ed}{\end{displaymath}}
\newcommand{\bi}{\begin{itemize}}
\newcommand{\ei}{\end{itemize}}
\newcommand{\bfig}{\begin{figure}}
\newcommand{\efig}{\end{figure}}
\newcommand{\bc}{\begin{center}}
\newcommand{\ec}{\end{center}}
\newcommand{\hii}{{H\scriptsize{II}}}

\newcommand{\vlsr}{V$_{\mathrm{LSR}}$}
\newcommand{\coa}{$^{12}\mathrm{CO}$}

\newcommand{\lsun}{L$_{\odot}$}
\newcommand{\msun}{M$_{\odot}$}
\newcommand{\lfir}{L$_{\textrm{\tiny{FIR}}}$}

\newcommand{\mcs}{M$_{\textrm{\tiny{CS}}}$}
\newcommand{\mdust}{M$_{\textrm{\tiny{dust}}}$}
\newcommand{\mvir}{M$_{\textrm{\tiny{virial}}}$}

\begin{document}


\title{Four Highly Luminous Massive Star Forming Regions in the Norma Spiral Arm II. Deep NIR imaging}


\author{L. Chavarr\'{\i}a\altaffilmark{1}, D. Mardones\altaffilmark{1,2}, G. Garay\altaffilmark{1}, A. Escala\altaffilmark{1}, L. Bronfman\altaffilmark{1} and S. Lizano\altaffilmark{2}}


\altaffiltext{1}{Departamento de Astronom\'{\i}a, Universidad de Chile,
    Camino del Observatorio 1515, Las Condes, Santiago, Chile}
\altaffiltext{2}{Universidad Nacional Autonoma de Mexico, Apt Postal 3-72 (Xangari), MX Morelia, Michoac\'an 58089, Mexico}


\begin{abstract}
We present sensitive NIR (J, H and K) imaging observations toward four luminous massive star forming regions in the Norma Spiral Arm: G324.201+0.119, G328.307+0.432, G329.337+0.147 and G330.949-0.174. We identify three clusters of young stellar objects (YSO) based on surface density diagnostics.  We also find that sources detected only in the H and K-bands and with colors corresponding to spectral types earlier than B2, are likely YSOs. We analyze the spatial distribution of stars of different masses and find signatures  in two clusters of primordial mass segregation which  can't be explained as due to incompleteness effects.  We show that dynamic interactions of cluster members with the dense gas from the parent core can explain the observed mass segregation, indicating that the gas plays an important role in the dynamics of young clusters. 
\end{abstract}





\section{Introduction}\label{section_introduction}
While still deeply embedded, massive star clusters keep an imprint of the physical conditions in the natal molecular cloud. Massive stars form in molecular cores with typical radii of $\sim 0.4$ pc, molecular number densities of $3\times 10^5$ cm$^{-3}$ and masses of $5\times 10^3$ M$_{\odot}$ \citep{plu97,fau04}.  Radio wavelength observations indicate that massive stars are formed in groups \citep{ho81,gar93} and that they are generally located in the center of massive and dense cores \citep{gar07};  an indication of mass segregation.    Thus, the study of young massive clusters is essential to fully understand the process of massive star formation.

Stellar mass segregation in a cluster can be either a consequence of dynamical evolution, or of biased massive star formation towards the core center. Dynamical mass segregation arises from kinetic energy equipartition between the cluster members. Massive stars will move slower than low-mass stars and sink to the center of the cluster in a process that takes place in a time comparable to the relaxation time (t$_r$) of the cluster \citep{bin87}. On the other hand, primordial mass segregation takes place much faster, in a massive star accretion timescale. During this process, massive stars are created in the center of the molecular cloud, where the gas density is highest \citep{bon01} and/or the mass accretion rate is high \citep{kru06}.  Several studies have shown evidence of mass segregation in young open clusters having ages of at least one order of magnitude smaller than their relaxation time \citep{gou04,bra05,che07}. This evidence suggests that mass segregation in those clusters is primordial. However, dynamical timescales are often calculated taking into account star-to-star interactions only, ignoring the gas. The later may have an impact in young clusters, where most of the mass is still in the form of gas.
Recent studies have shown that the gaseous component may have an important effect in the dynamics of the stars by gravitational drag, making dynamic times even shorter \citep[][in \S~5.2]{esc03,esc04}. Thus, studies of mass segregation in young clusters should
include the effects of dense gas in the cluster dynamics. On the other hand, Ascenso et al. (2009) have argued that sample incompleteness in observational data can mislead mass distribution analysis, producing mass segregation in non-segregated clusters. Therefore, to assess whether or not there is mass segregation in young clusters it is necessary to use robust mass segregation indicators.

In this paper we study the spatial distribution of stars in four luminous massive star forming regions located in the Norma spiral arm, at distances between 5.4-7.3 kpc from the Sun. The regions were observed in the near-infrared J, H, and K bands at the VLT.   Garay et al. (2009, hereafter Paper I) present dust mm-continuum observations towards these same regions.   Target selection and data reduction are presented in \S~\ref{section_observations1}. Results, including identification of young stars, analysis of their spatial distribution and extinction maps are presented in \S~\ref{section_results1}. In \S~\ref{section_discussion1} we study the radial distribution of cluster members of different masses. We also discuss the age of the clusters, the evidence of mass segregation and the presence of gaps in the color-magnitude diagrams. Conclusions are presented in \S~\ref{section_conclusions1}.

\section{Observations and data reduction}\label{section_observations1}
\subsection{Source Selection}
We observed the regions G324.201+0.119, G328.307+0.432, G329.337+0.147 and 
G330.949-0.174, selected from the CS(2$\rightarrow$1) survey of \citet{bro96} 
by being among the brightest objects at far-infrared (FIR) wavelengths 
(\lfir$ > 6\times10^5$~\lsun). The four regions are located in the tangent of the 
Norma spiral arm (see Table~\ref{sources}) and their CS(2$\rightarrow$1) line profiles 
show broad line-widths (FWHM between 6.1 and 6.9~km~s$^{-1}$) and the presence 
of wings, making them good candidates for young clusters containing massive stars.

The LSR velocities of G324.201 and G329.337 are close to the terminal velocities, which are the maximum allowed by pure circular motion along their lines of sight, thus,  we locate them at the sub-central distances.  These two objects are located well away from the centers of their parental molecular clouds as traced by CO \citep{bro88}.  We assign the near kinematic distances based on the distance of the molecular clouds in which they are deeply embedded \citep{bro00} to the sources G328.307 and G330.949.   These sources represent massive star forming regions in different environments, i.e., in the center and in the periphery of giant molecular clouds.

\subsection{Near-IR data}\label{nir_data}
The four sources were observed with ISAAC at the VLT-Antu in April 1999, using an $1024\times 1024$ pixel InSb Aladdin detector array with a scale of 0.147 arcsecond per pixel. The typical seeing during the observations was 0.4 arcseconds. We observed in the J, H and K bands centered at 1.24 1.65 and 2.16 $\mu m$ respectively. The observations were done in dithering mode with 30 arcsecond shifts. The integration times are shown in Table~\ref{time}.

The data were corrected for bias, dark current, and pixel to pixel sensitivity variations. The XDIMSUN task in IRAF was used to create a sky image using the median time average of the dithered observations. We subtracted the sky image and corrected the field distortion\footnote{http://www.eso.org/instruments/isaac/field\_distortion.html} by transforming the individual images to a central reference image using second order transformations in X and Y as recommended by ESO. This had a minor impact ($< 1"$) in the source position at the edges of the detector. The field distortion correction, however, improved the roundness of the stellar images increasing the photometric precision. Finally, we median combined the resulting images and defined the coordinate system using WCSTools\footnote{http://tdc-www.harvard.edu/software/wcstools/} and the 2MASS catalogue.

\subsection{Photometry}\label{photometry}
We searched for star candidates using the DAOPHOT package in IRAF with a  3 sigma detection threshold. Then we performed aperture photometry using an aperture radii of 0.45 arcseconds. The typical full width at half maximum (FWHM) of the detected sources was 2.7 pixels or 0.4 arcseconds. Sky emission was eliminated by subtracting the median value inside a ring with inner radius of 1.5 arcseconds and 1.5 arcseconds of width. Finally, a carefully visual inspection was done over the whole mosaic and in particular where there was extended emission in order to reject diffuse features and saturated stars. 

\begin{figure}
\begin{center}
\epsscale{0.6}
\plotone{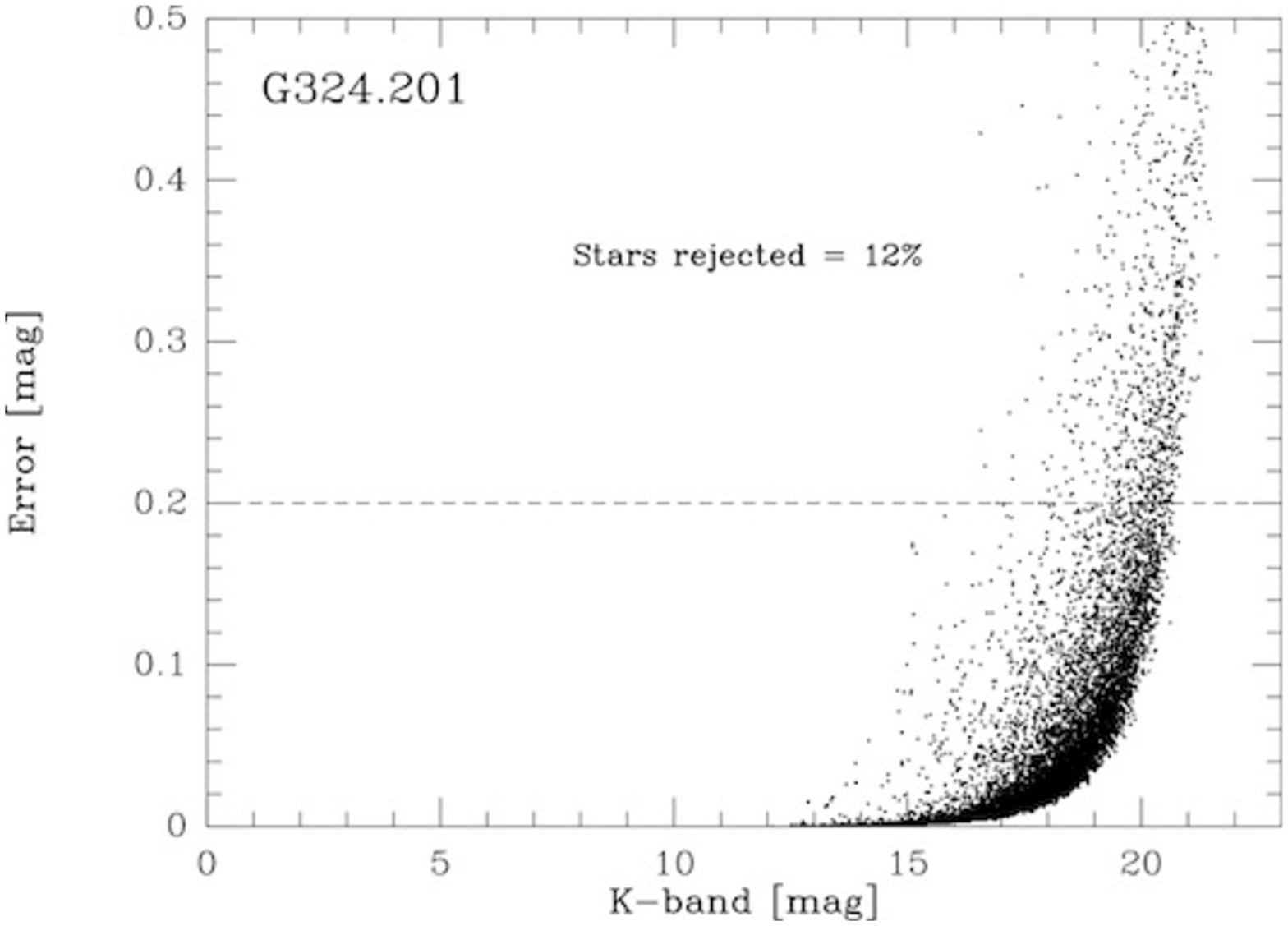}
\plotone{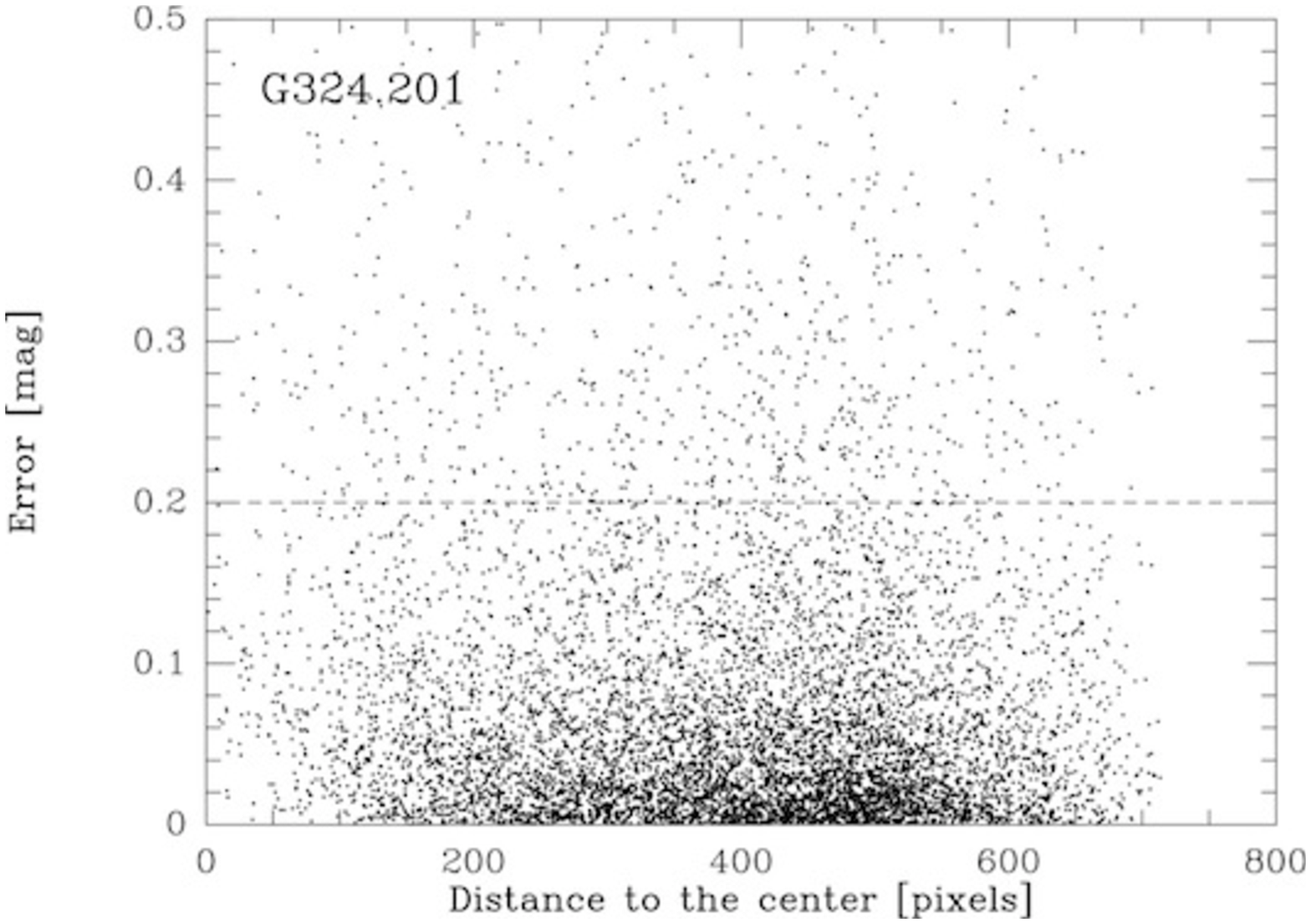}
\caption{Errors in the K-band photometry for G324.201. \emph{Upper panel}: error as a function of K-band magnitude. \emph{Lower panel}: error as a function of the distance to the center of the mosaic.  The segmented line indicates the maximum error allowed. The error is independent of the distance to the center of the mosaic.\label{err}}
\end{center}
\end{figure}

Calibration was performed by minimizing residuals to corresponding 2MASS detections. The RMS for residuals between the data sets used was less than 0.05 magnitudes in all filters. Finally, we combined J, H, and K photometry by merging the data. A source in different bands was identified as the same if the difference in their position was less than 0.3 arcseconds. Figure~\ref{err} shows the photometric error dependence over magnitude (upper panel) and over distance from the center of the mosaic in the K-band for G324.201 (lower panel). We only use sources with an error of less than 0.2~mag in all bands. This corresponds to 88\% of the detected sources in G324.201 and G328.307, and to 77\% in G329.337 and G330.949. Sources with an error of more than 0.2~mag are distributed uniformly over the mosaic.

\section{Results}\label{section_results1}
\begin{figure}
\begin{center}
\epsscale{1.1}
\plottwo{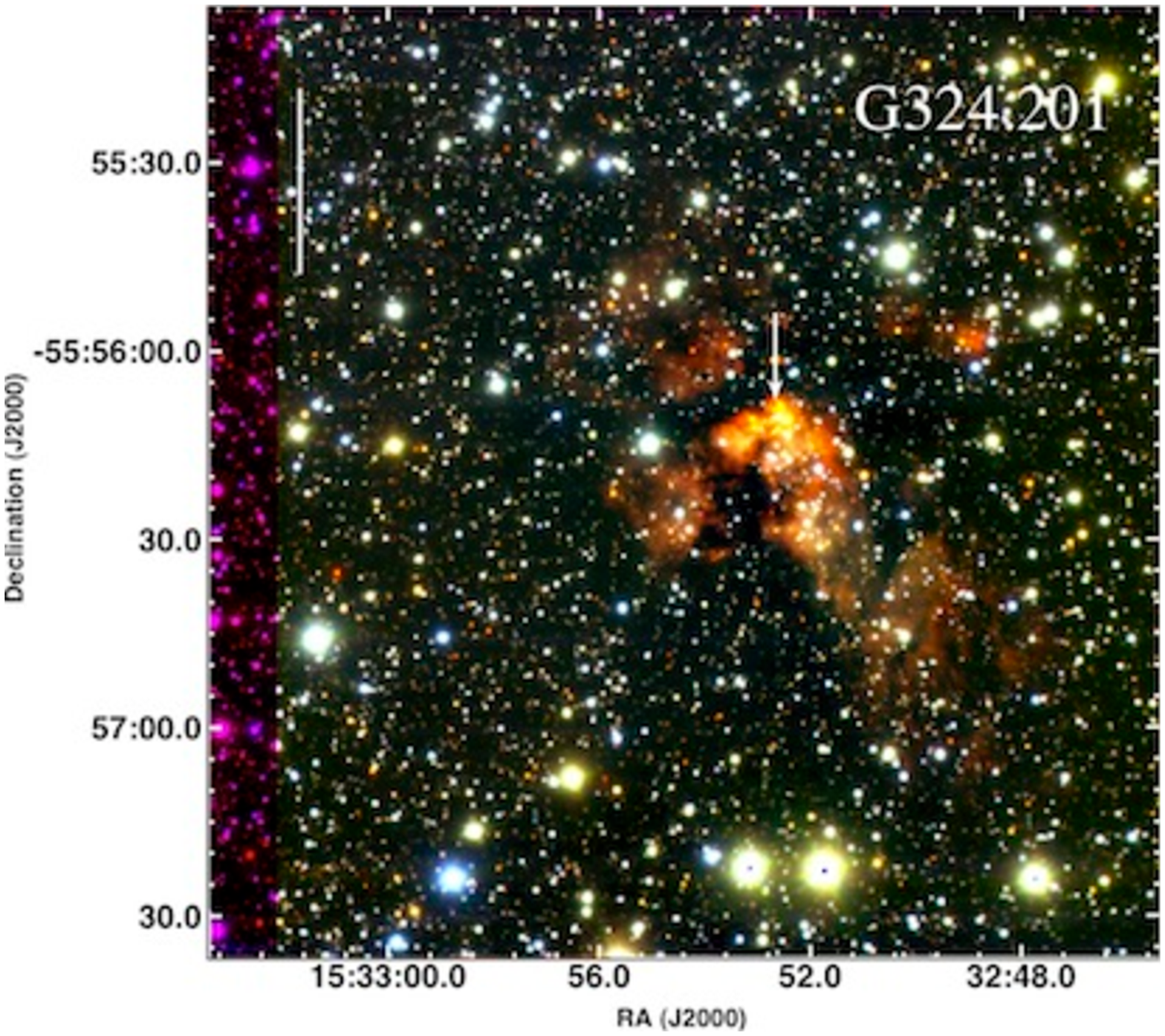}{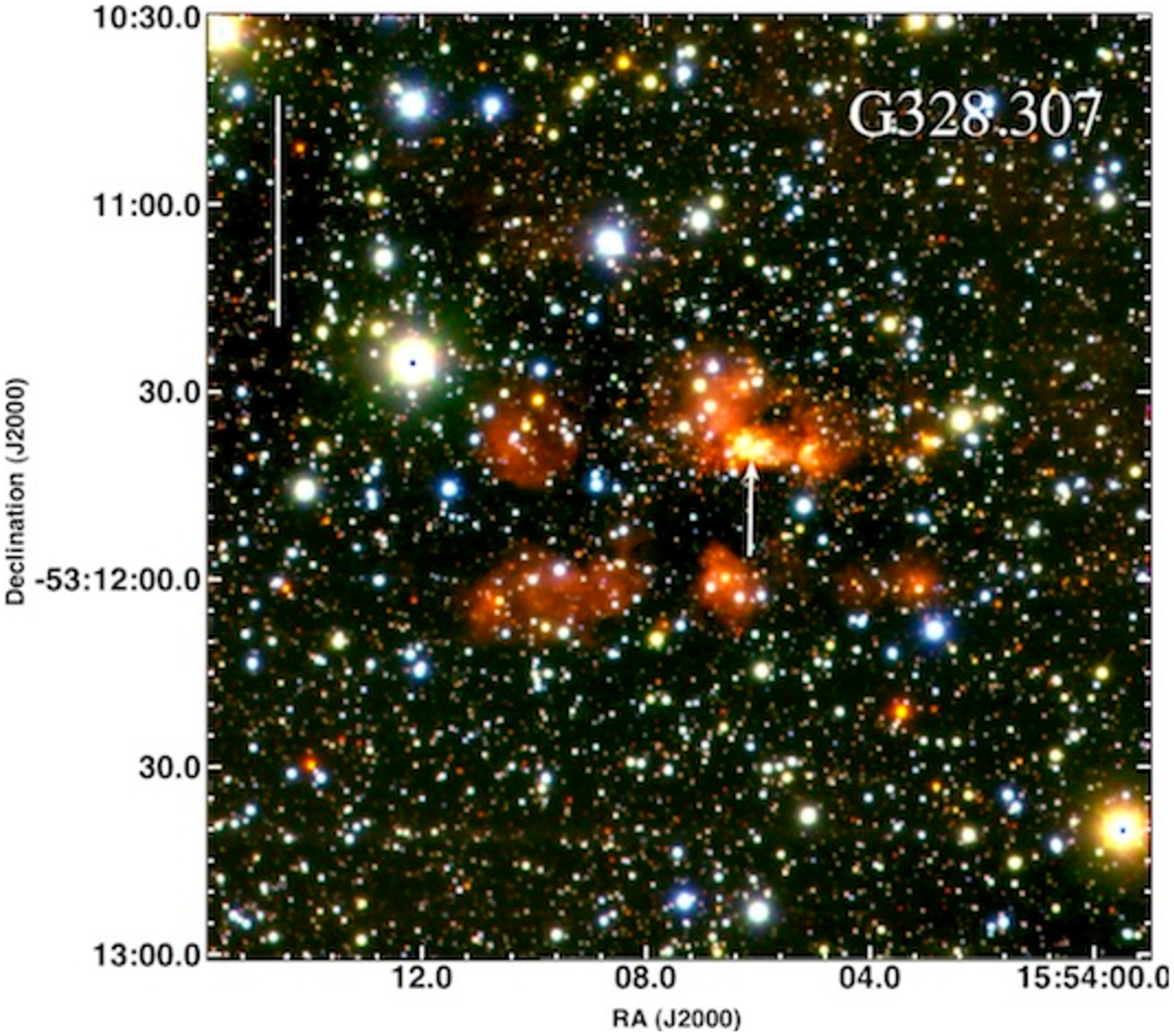}
\plottwo{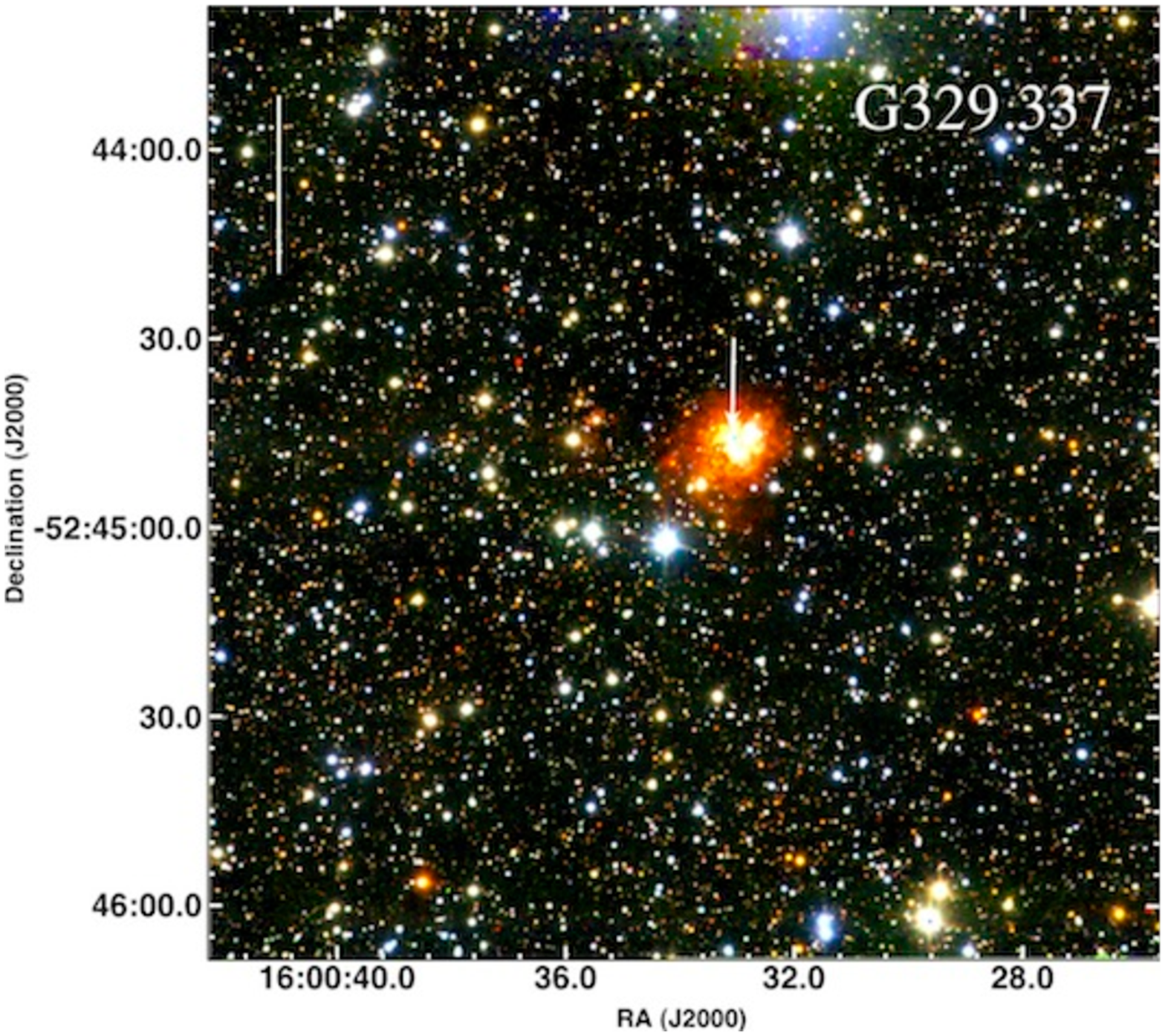}{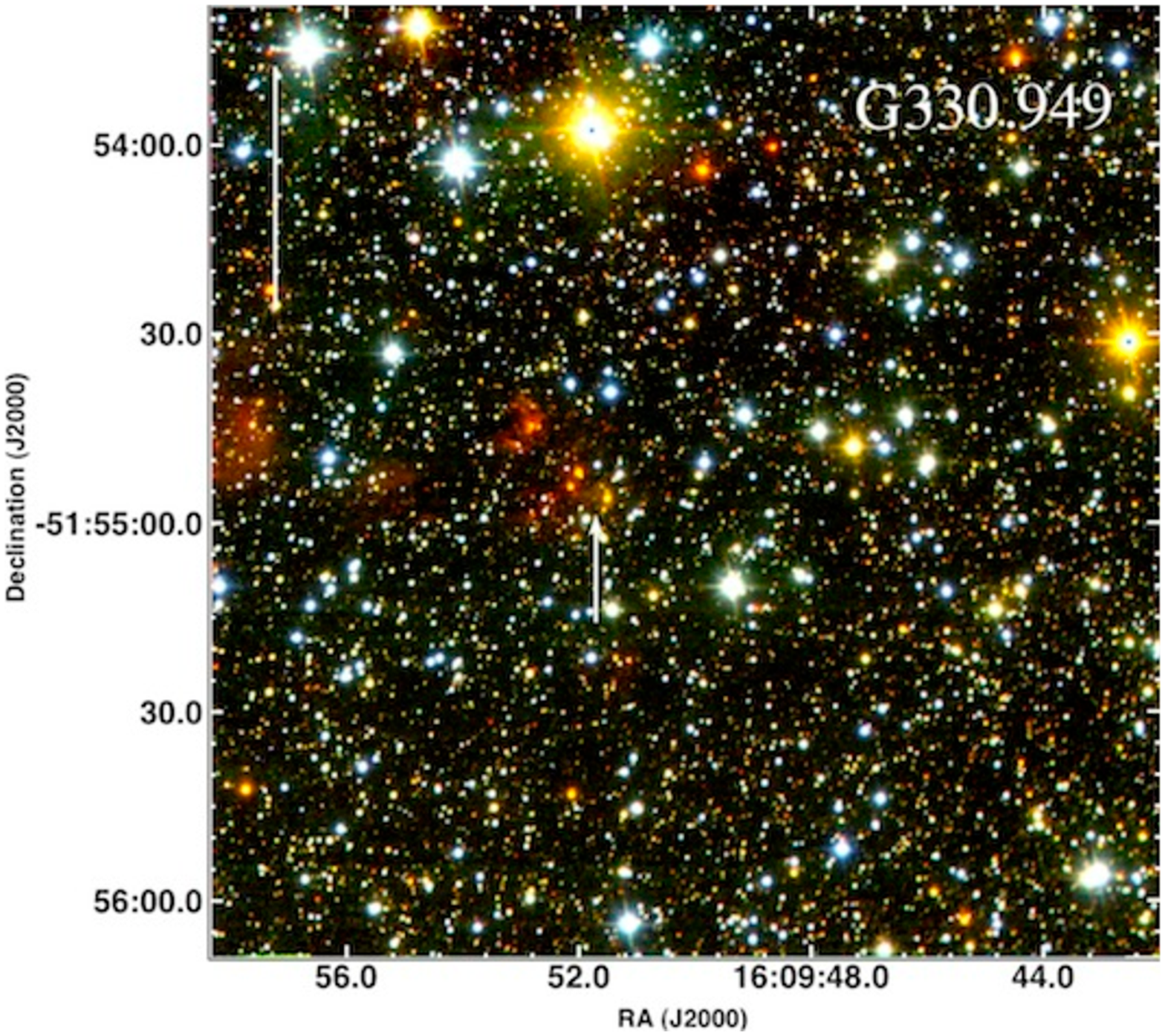}
\caption{JHK-color image of the regions observed (J-blue, H-green, K-red). The white bar represents 1 parsec. The arrow indicates the position of the center of the clusters.\label{JHK}}
\end{center}
\end{figure}

\begin{figure}
\begin{center}
\epsscale{0.8}
\plotone{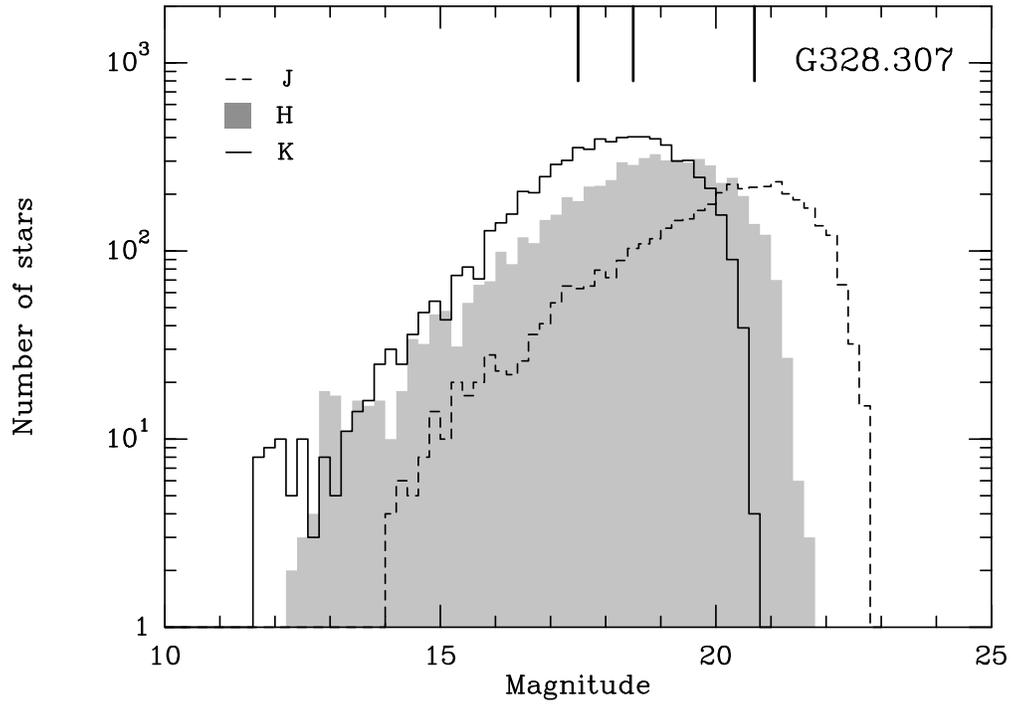}
\caption{Apparent magnitude histogram for detections with photometric errors $< 0.2$ magnitudes in all bands for G328.307.   The vertical lines on the top indicate the 90\% completeness limits in the J, H and K bands: 20.7, 18.5 and 17.5~mag respectively.\label{jhk_histo}}
\end{center}
\end{figure}

Figure~\ref{JHK} shows composite three-color images of the four regions observed. In G324.201 the field of view is not fully covered in the H-band.  Extended emission is clearly seen in all regions with the exception of G330.949 and shows a more complex structure in G324.201 and G328.307 than in  G329.337. Since this emission is seen in the J and K-bands, we hypothesize that most of it corresponds to stellar emission reflected by the dust. Table~\ref{stars} lists the total number of stars detected in each of the bands with error $< 0.2$ mag. The completeness of our data (shown in Table~\ref{completeness}) corresponds to the magnitude at which at least 90\% of the stars is detected and it was estimated by the fraction of artificial stars recovered.  Figure~\ref{jhk_histo} shows the histogram of the detections in each band for G328.307.

\subsection{Color-color diagrams}\label{redal}
\begin{figure}
\begin{center}
\epsscale{0.9}
\plottwo{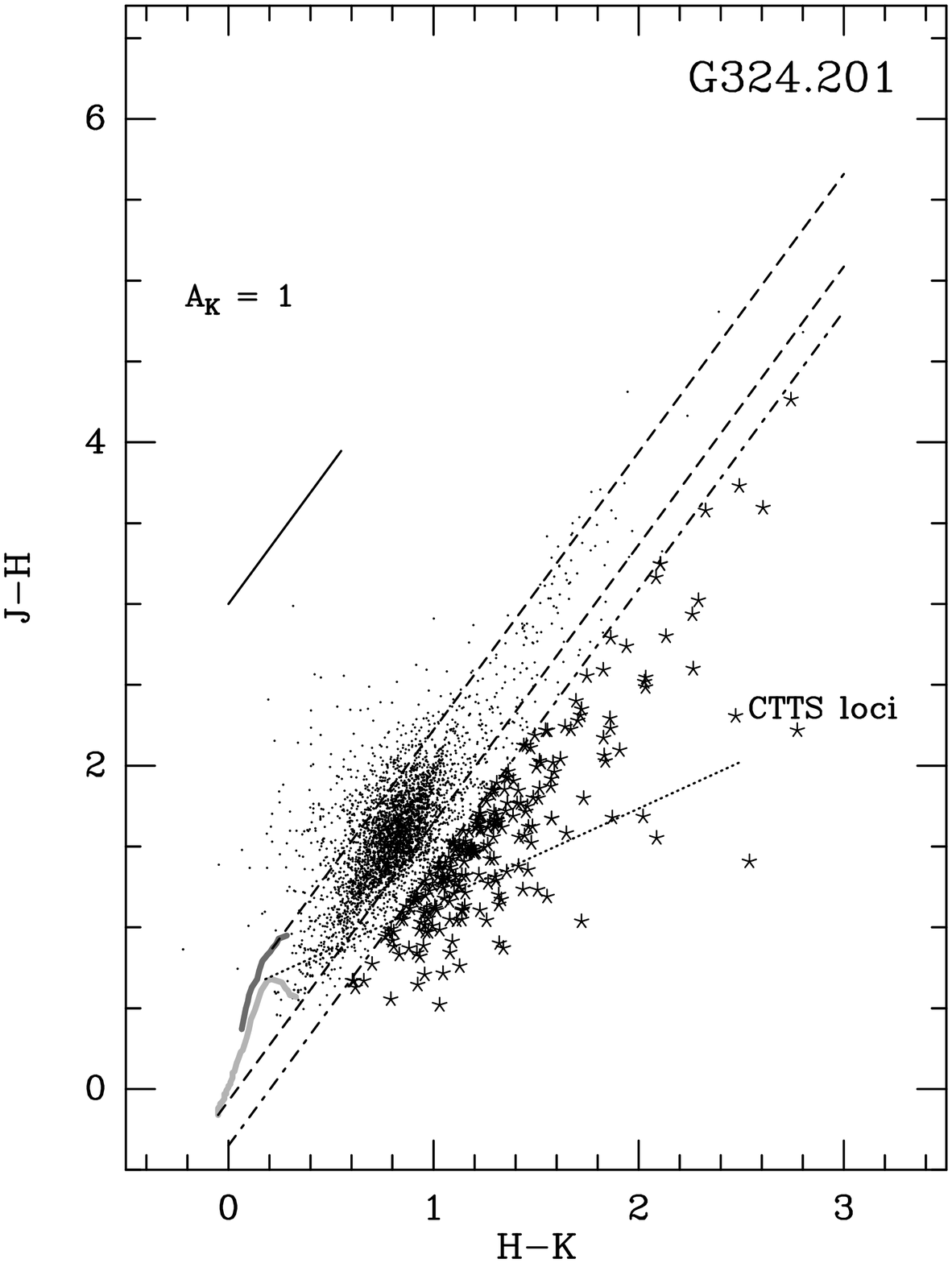}{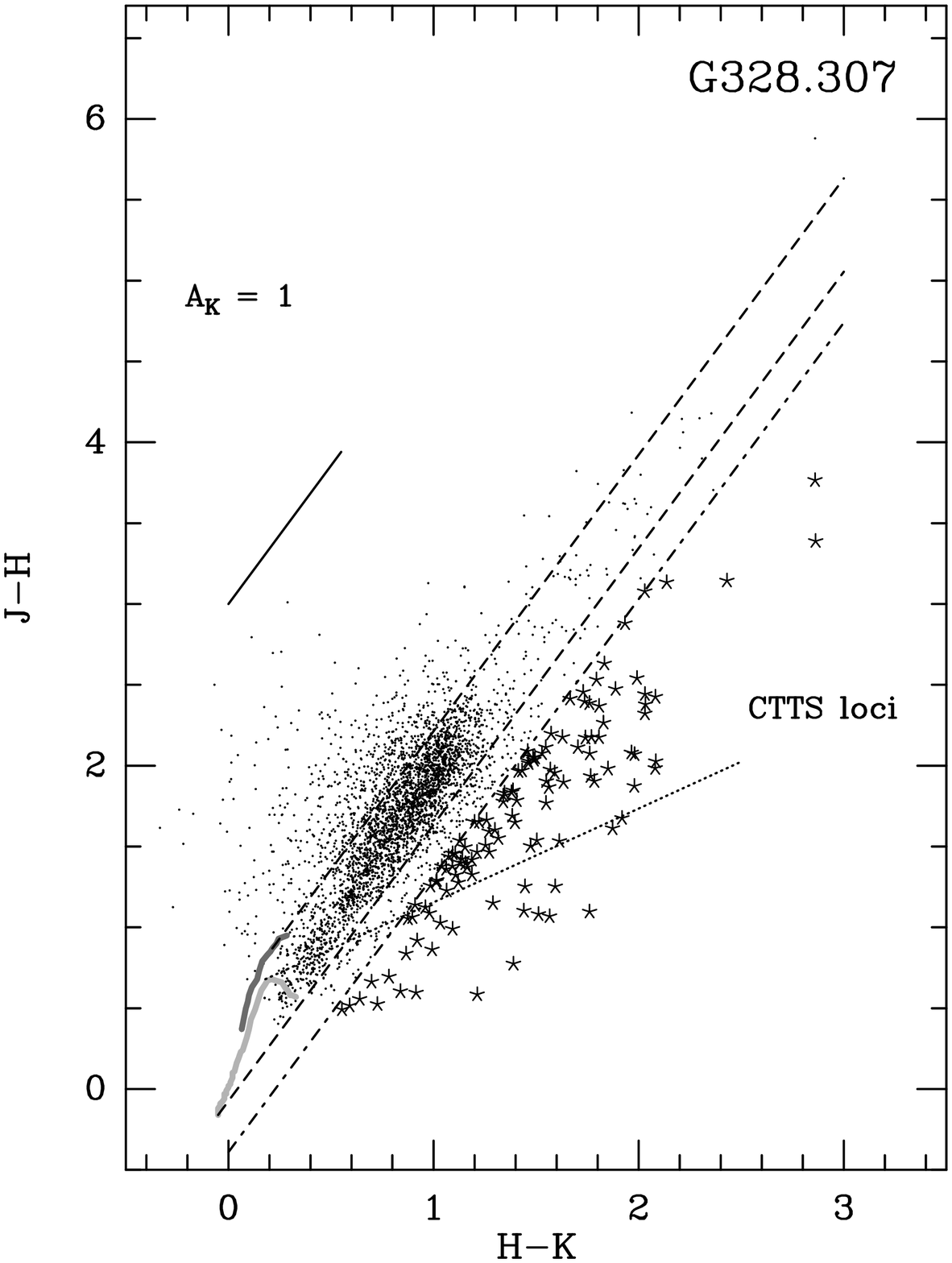}
\plottwo{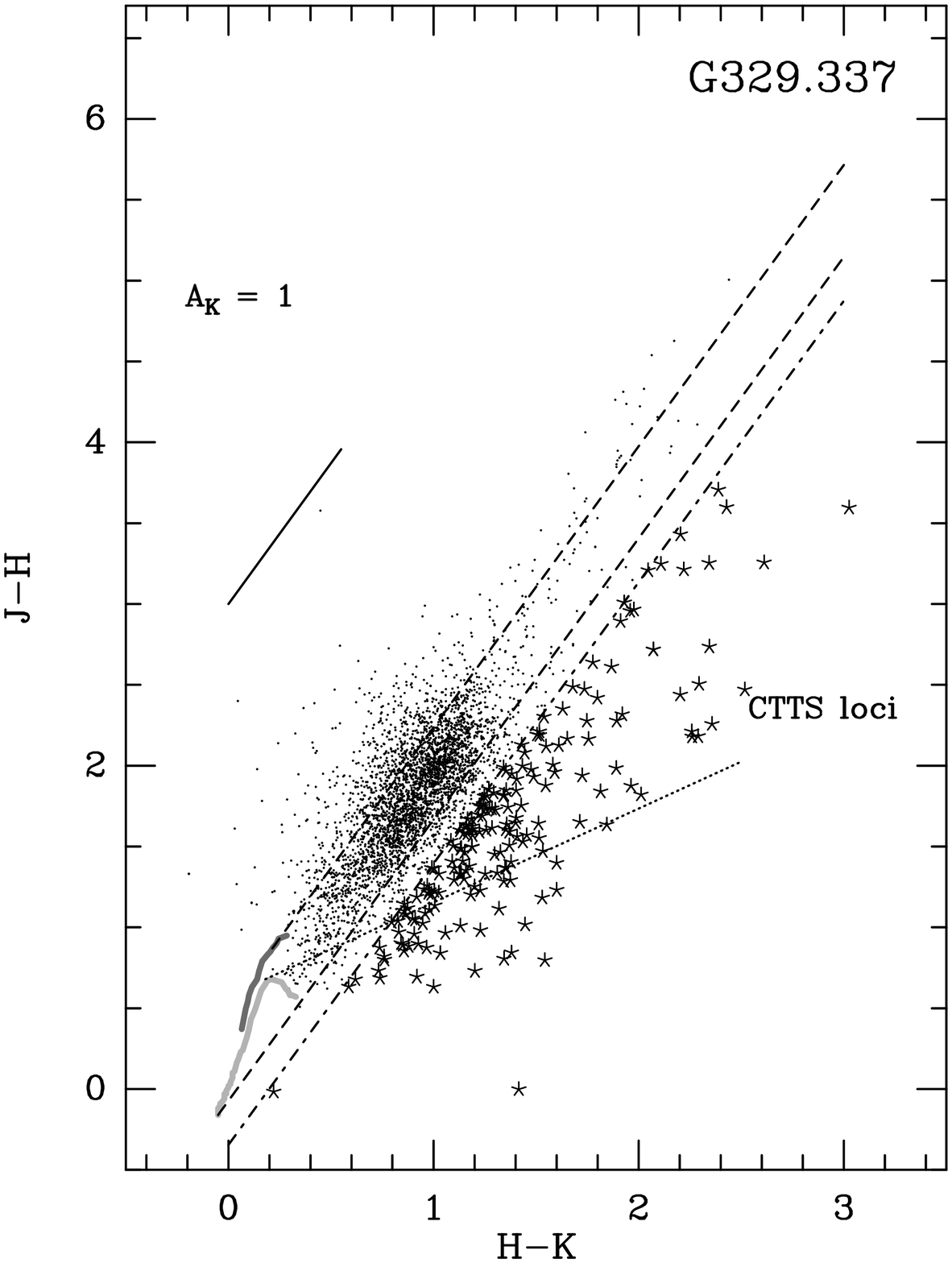}{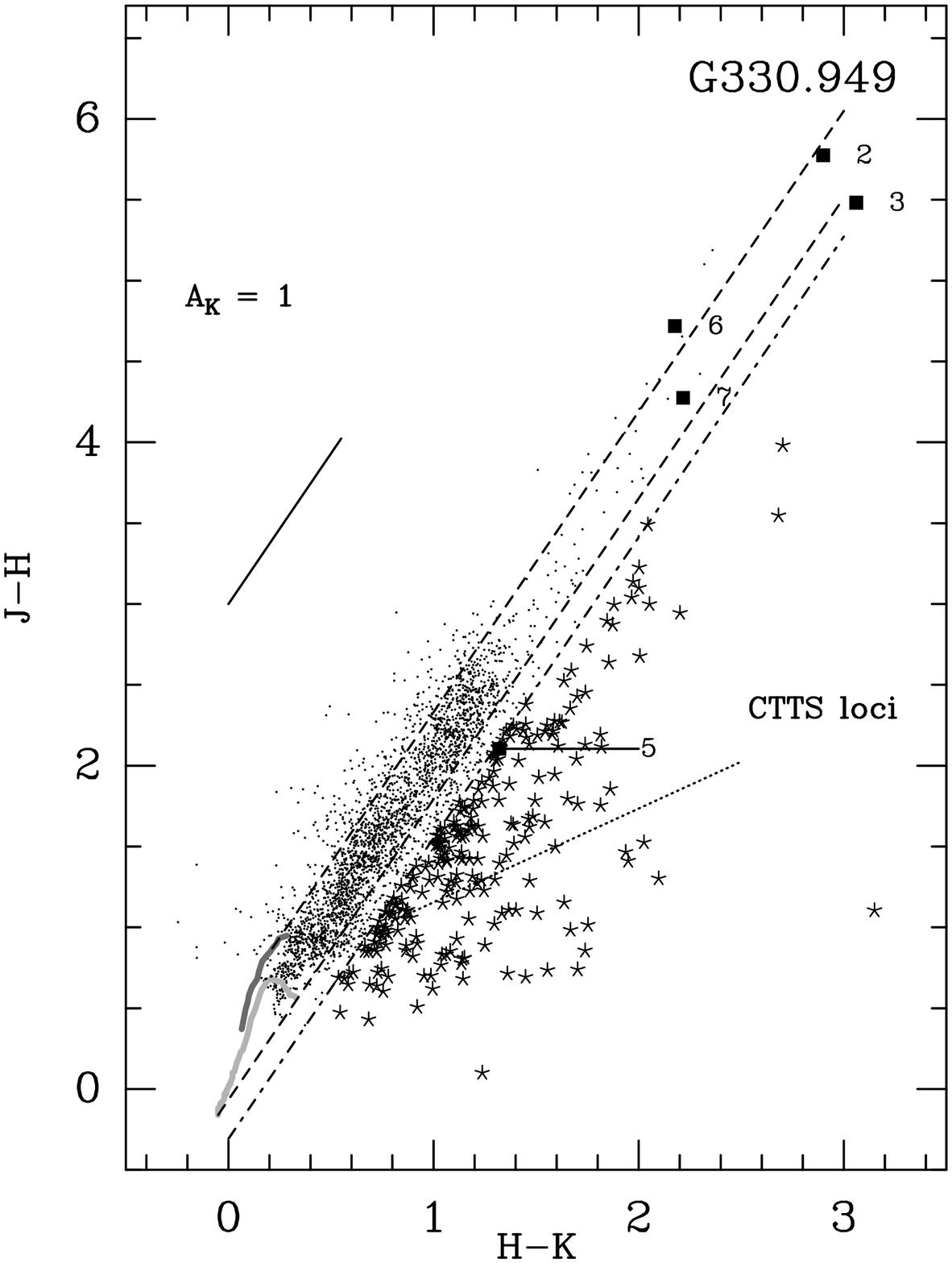}
\caption{JHK color-color diagrams for the four regions. Gray lines correspond to the position of the main sequence (light gray) and giant stars. Dashed lines enclose reddened main sequence and giant  stars.  The dashed-dotted lines are displaced by the $\Delta$ dispersion of the sources located above the giant branch.  Sources with NIR-excess are marked as asterisks.  The dotted lines indicate the CTTS loci from \citet{mey97}. The percentage of sources located above the giant branch is 3.5, 10, 6 and 2.5\% for G324.201, G328.307, G329.337 and G330.949 respectively. Gray squares in G330.949 correspond to individual sources from Figure~\ref{g330} (see \S~\ref{section_g330}).\label{colcol}}
\end{center}
\end{figure}

Color-color diagrams of the stars within the observed regions are shown in Figure~\ref{colcol}. Most of the stars are located in the region between the reddening vectors and they correspond to normally reddened stars. A small percentage of stars (between 2.5 and 10\%) are distributed above the main sequence. This is typically seen in other regions of star formation, \citep[eg. M17, Westerlund 2 and NGC7538, from][respectively]{jia02,asc07,bal04}.  We found that these sources are distributed uniformly over the field and do not have a characteristic brightness. The scatter in the color-color diagram was quantified by computing the mean distance to the giant branch, $\Delta$, of the sources located above it.   This scatter is not due to systemic errors in the reduction of the data and affects a small percentage of sources. 

We consider sources with NIR-excess those located at least one $\Delta$ to the right of the reddening vector (marked as asterisks in Fig.~\ref{colcol}). Some of these sources may be background galaxies. To estimate the number of background contaminants we used the K-band galaxy number count from \citet{bla08} and found a maximum of 10 galaxies in an area of the same size as the area covered by our data (with K $< 18$ mag). Since the number of galaxies in the FOV is small, we assume that all the NIR-excess sources correspond to young stellar objects. We found 236, 120, 186 and 228 sources with NIR-excess in G324.201, G328.307, G329.337 and G330.949, respectively.

\subsubsection{Extinction law}\label{section_extinctionlaw}
\begin{figure}
\begin{center}
\epsscale{0.9}
\plottwo{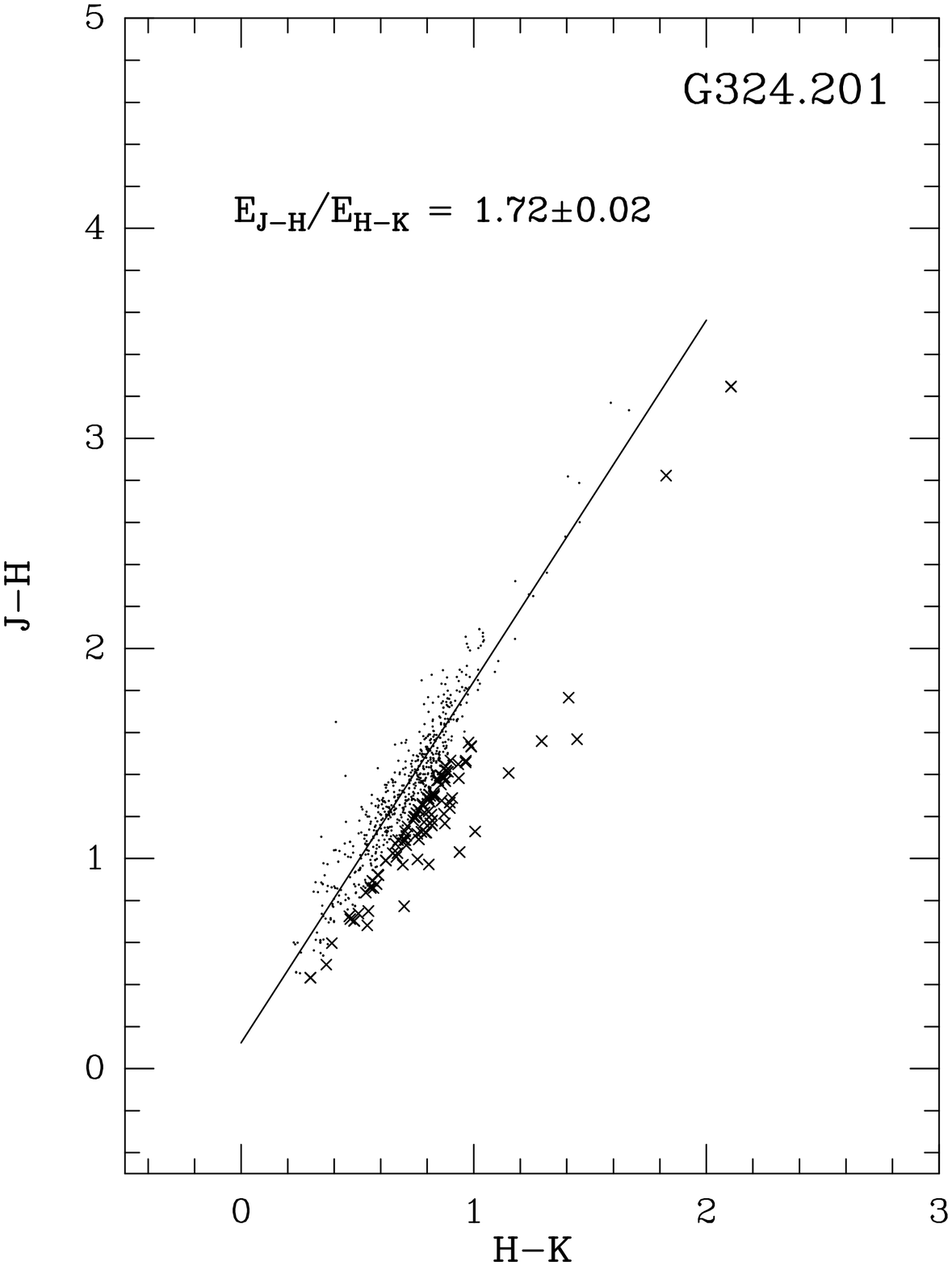}{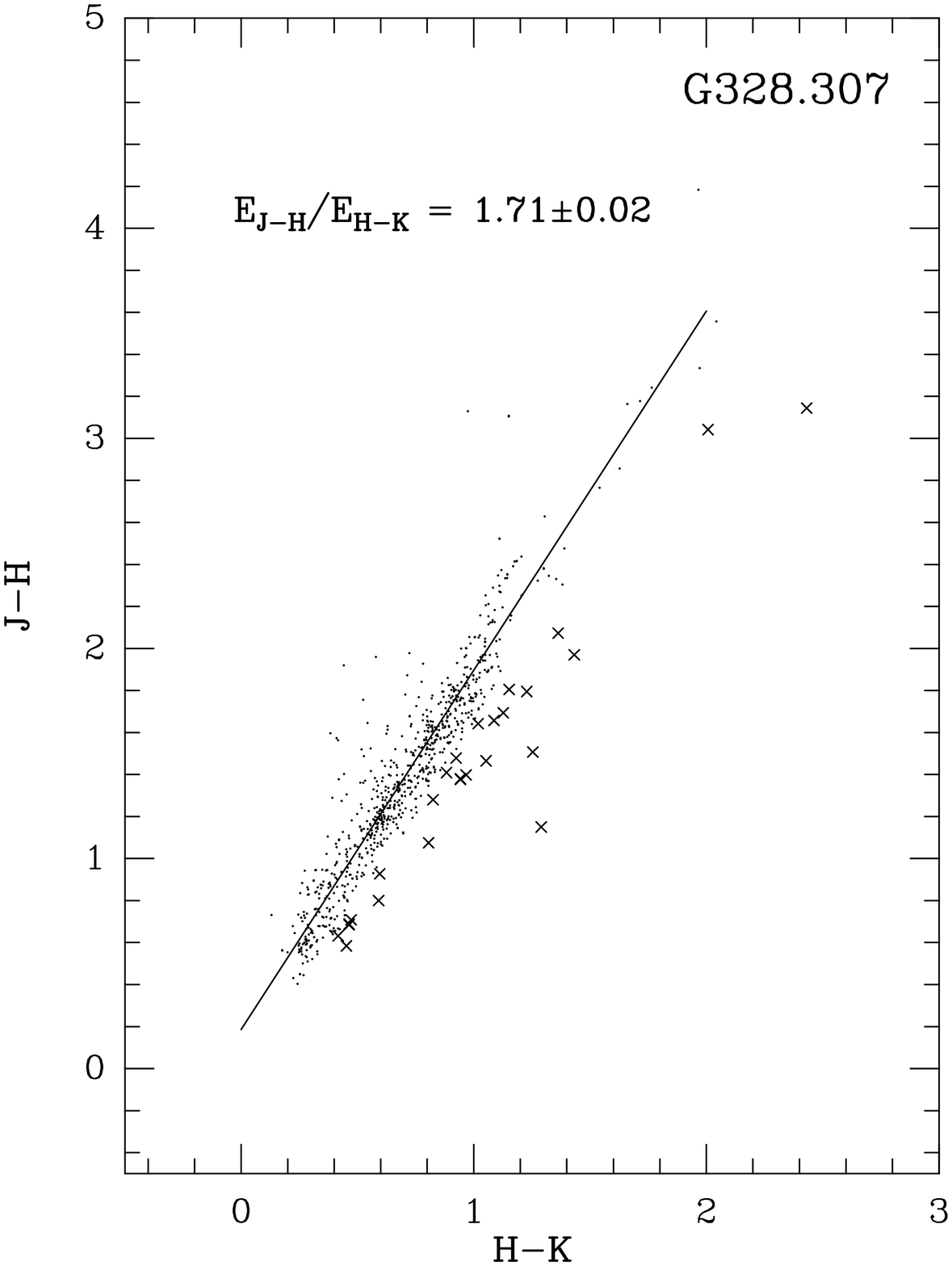}
\plottwo{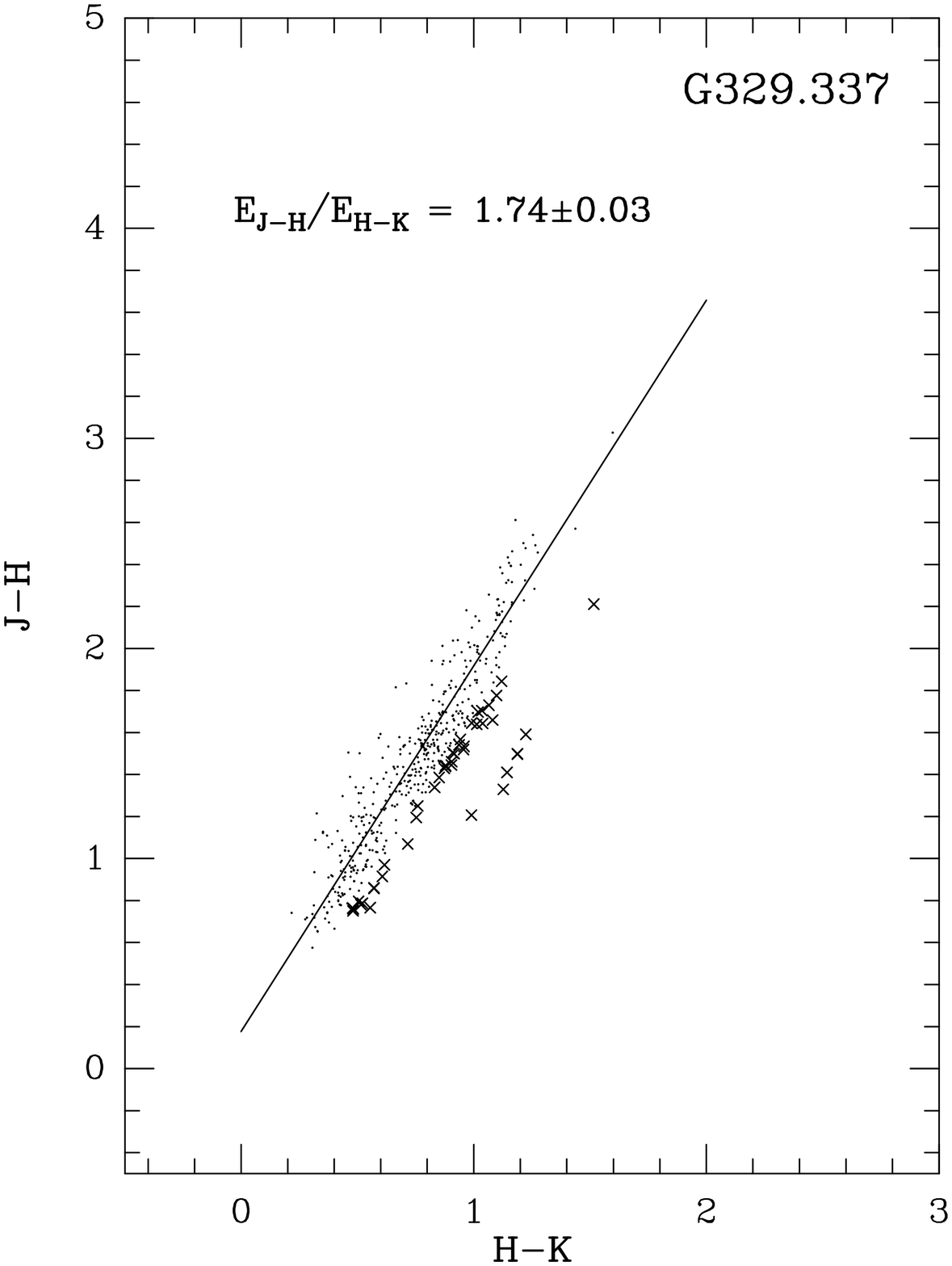}{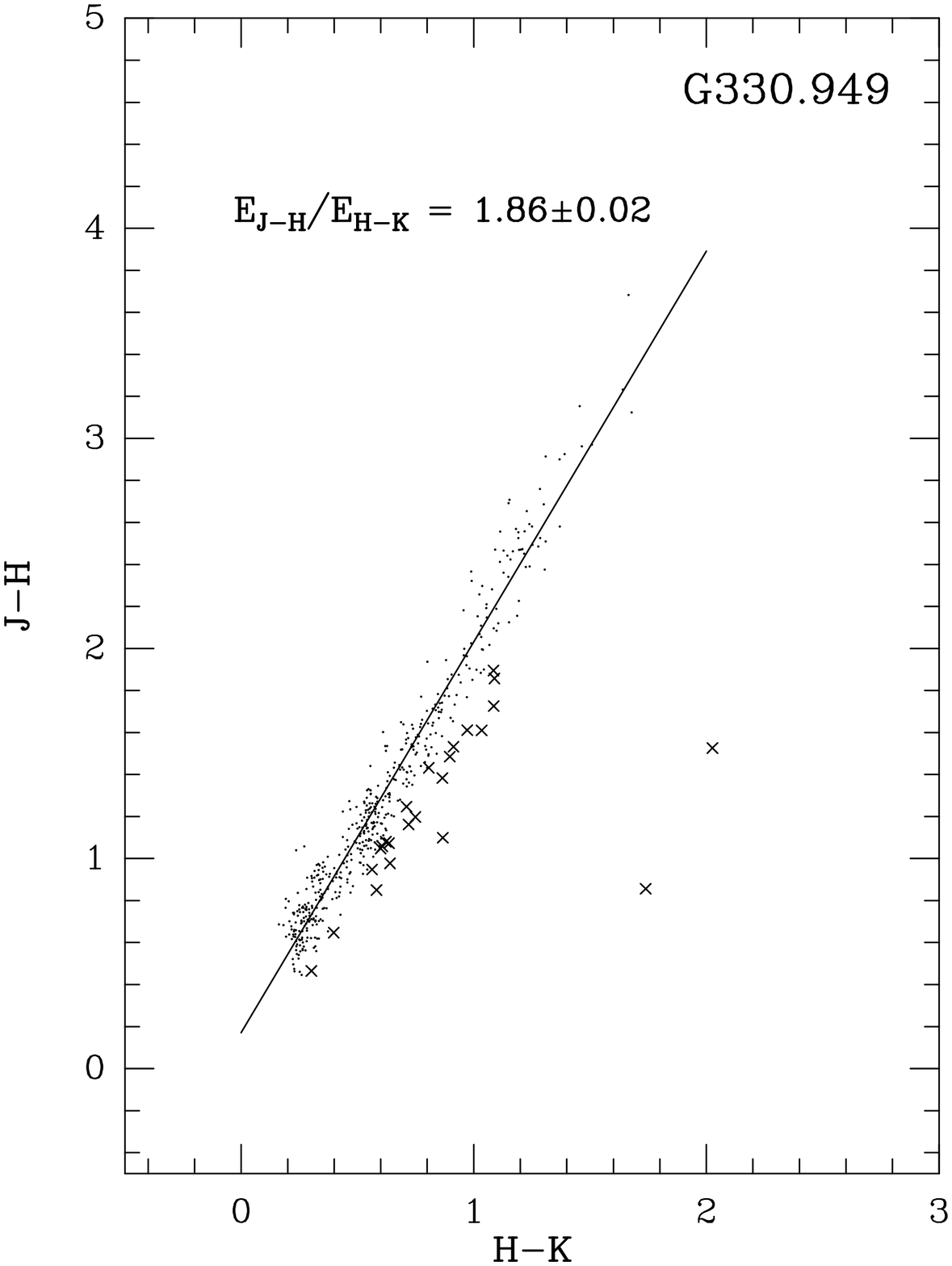}
\caption{JHK color-color diagram used to determine the extinction law in each region.  We used stars with photometric errors smaller  than 0.01 magnitudes in all bands. Points marked as "x" correspond to stars eliminated by the algorithm explained in~\S~\ref{section_extinctionlaw}.  The solid lines indicate the extinction law found in each region.\label{reddening}}
\end{center}
\end{figure}

We used the NIR color-color diagrams to determine the extinction law $E_{J-H}/E_{H-K}$ in each region. In order to do this, we eliminated NIR-excess sources from the sample using the following algorithm. First, we fit a vector to the data (a straight line in the color-color diagram) using all the stars with photometric errors of less than 0.01 mag. This vector will have a flatter value of $E_{J-H}/E_{H-K}$ than the true reddening vector since stars with excess are located below it. Then, we remove all the stars having NIR-excess with respect to this vector and fit a new vector to the remaining stars.  We repeat this process until the value of $E_{J-H}/E_{H-K}$ converges.  Figure~\ref{reddening} shows the resulting reddening vectors for the four regions. The values of $E_{J-H}/E_{H-K}$ range between 1.71 and 1.86 and are in agreement with previous values from the literature \citep[$E_{J-H}/E_{H-K}=1.7$ to $2.1$,][]{ken98,jia02}. 

\subsection{Color-magnitude diagrams}\label{section_colmag}
\begin{figure}
\begin{center}
\epsscale{0.9}
\plottwo{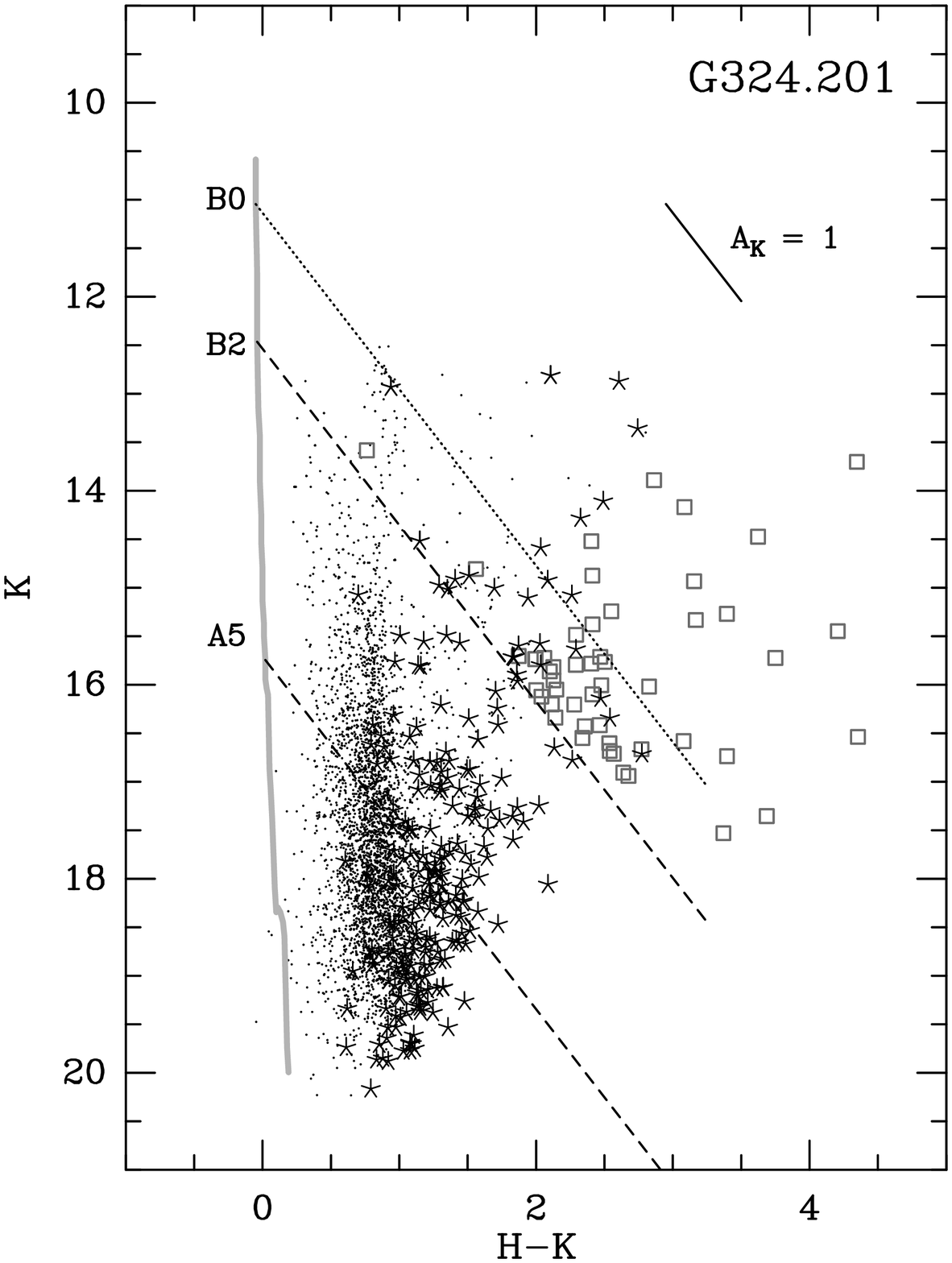}{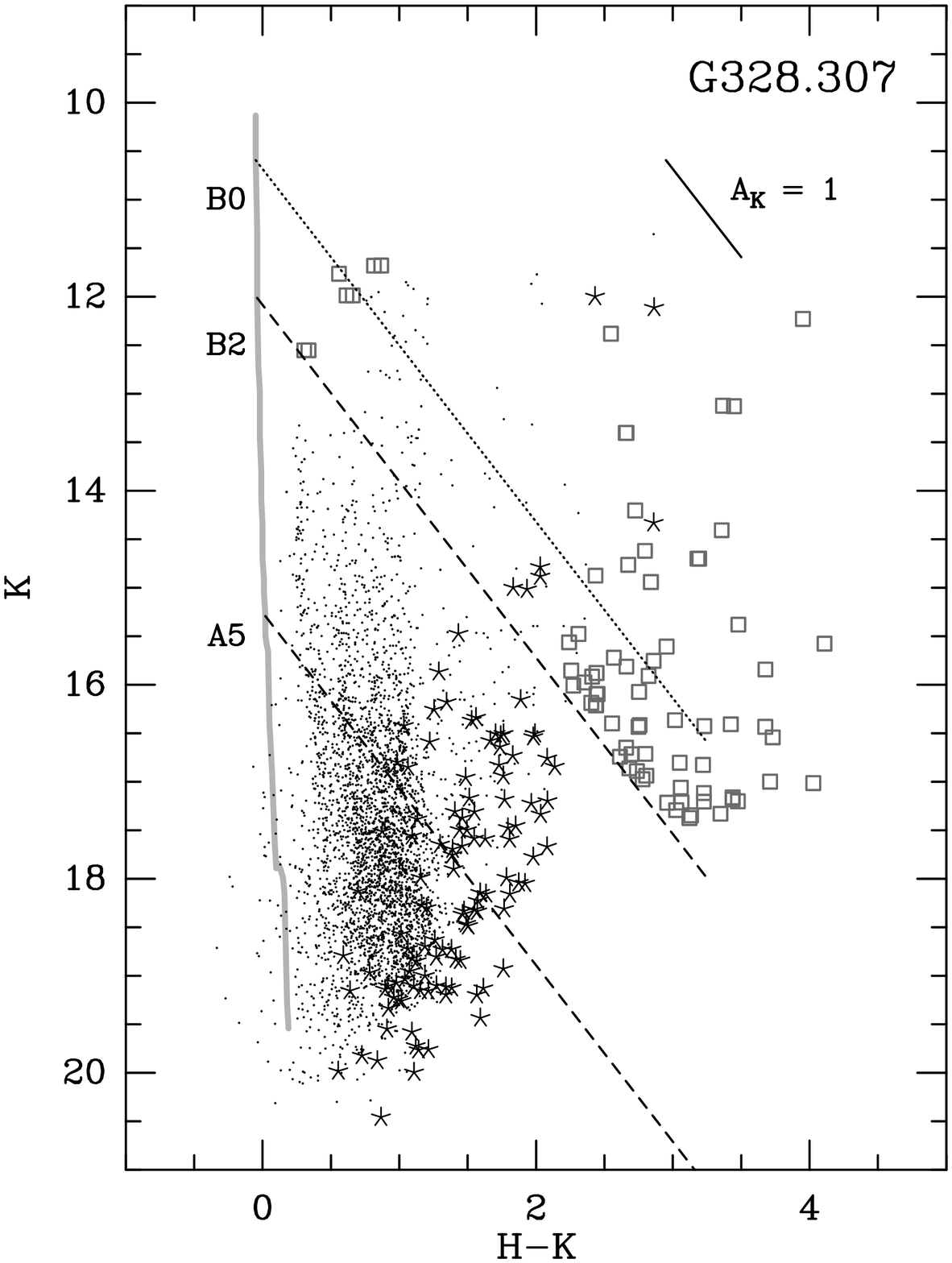}
\plottwo{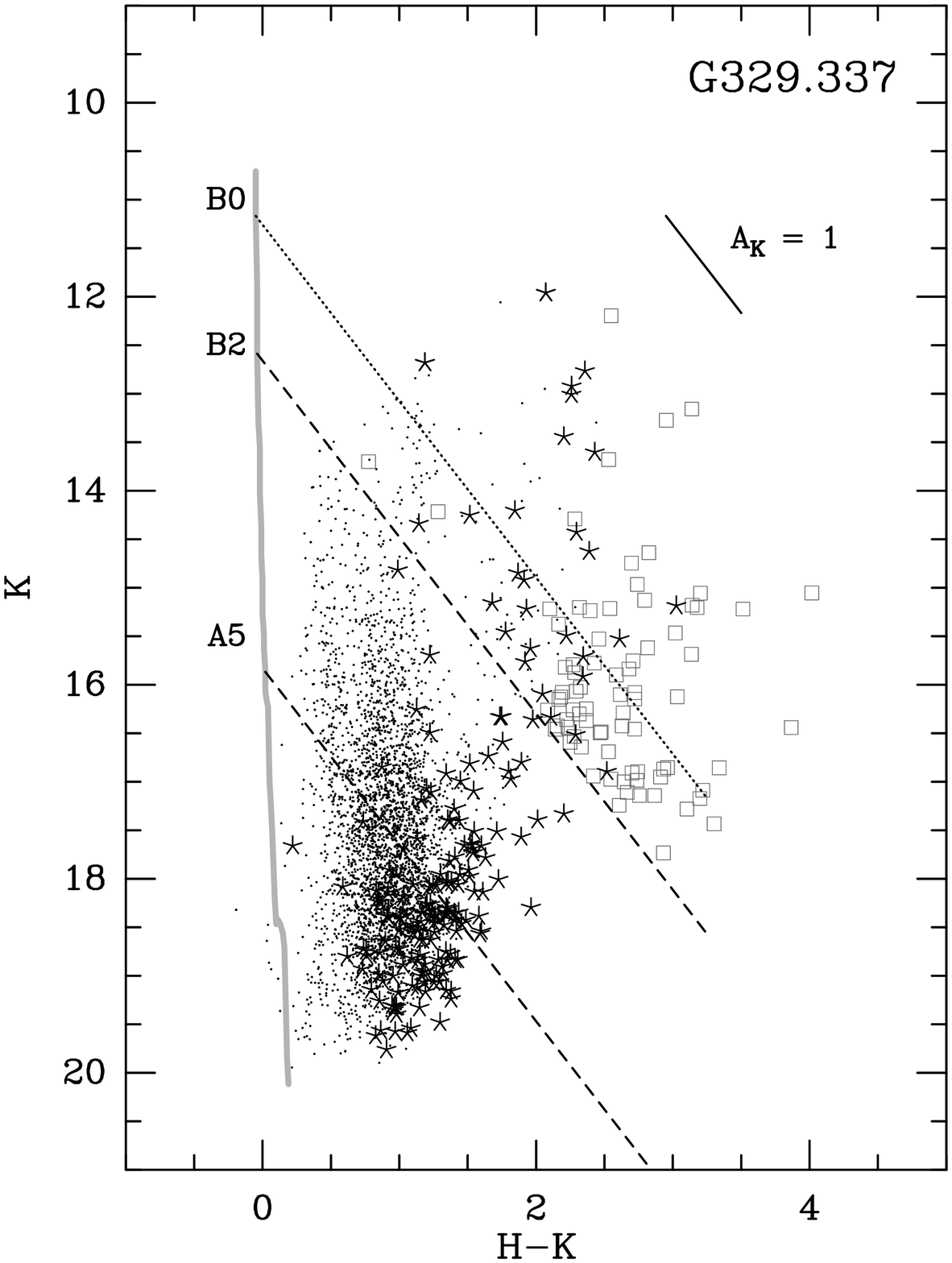}{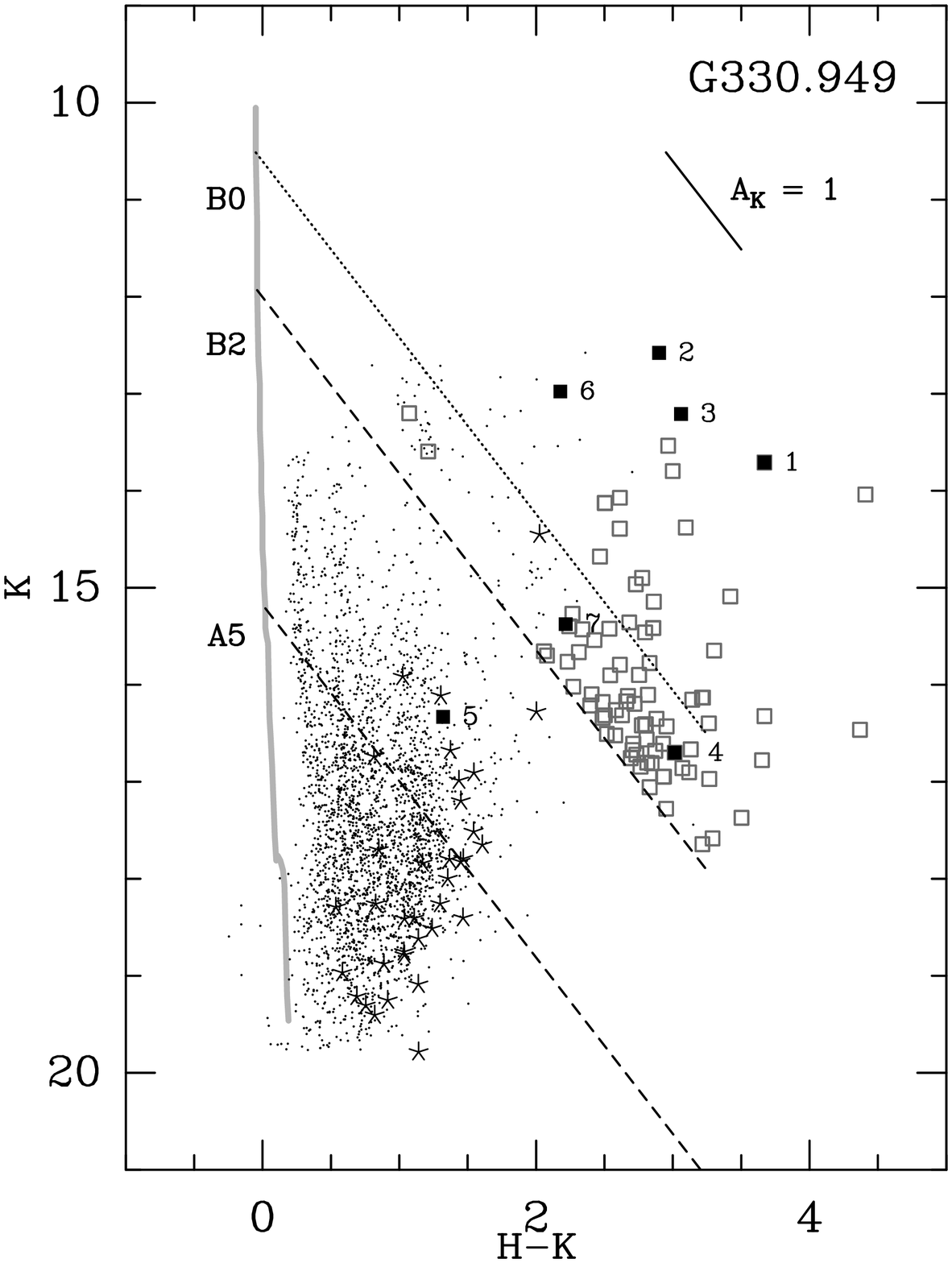}
\caption{Color-magnitude diagrams for the regions observed. The gray lines show the position of the main sequence stars \citep{koo83} at the adopted distances. Dashed lines correspond to the reddening vector for B0V, B2V and A5V stars. Sources with NIR-excess are marked as asterisks. Selected J-dropout sources are marked as empty squares. The diagrams show that the color of the stars is shifted from the main sequence due to the interstellar extinction. It is possible to identify several main sequences with different foreground extinction in each region. Black squares in G330.949 correspond to the individual sources from Figure~\ref{g330} (see \S~\ref{section_g330}).\label{color_magnitude}}
\end{center}
\end{figure}

Color-magnitude diagrams for all star forming regions are presented in Figure~\ref{color_magnitude}. NIR-excess sources are marked with asterisks. We used the color-magnitude diagram to divide the population of stars in two groups: \emph{massive stars} (stars with colors corresponding to spectral type earlier than B2) and \emph{intermediate-mass} stars (with colors corresponding to spectral types between B2 and A5). The magnitude of an unreddened A5 star located at the same distance than G329.337 (the furthest region) is K=16~mag, significantly brighter than our completeness limit (see Table~\ref{completeness}) indicating that we are complete in detections for both groups. The total number of  stars earlier than A5 detected in each region is listed in Table~\ref{stars}. 

Several sources were detected only in the H and K bands (see Table~\ref{stars}). These J-dropout sources are likely the most embedded YSOs.  However, it is difficult to distinguish YSOs from main sequence stars among the J-dropout sources because  in this case the color-color diagnostic used to find NIR-excess sources cannot be used.   Instead, we selected J-dropout sources that are brighter than a B2 type star. Since the K-band is more sensitive to IR-excess, YSOs will be preferentially brighter in this band. The location of the selected J-dropout sources in the H-K vs. K color-magnitude diagram are indicated by the squares in Figure~\ref{color_magnitude}. Their spatial distribution is discussed in \S~\ref{section_spatial_Jdropout}.

\subsection{Surface density of NIR-excess sources}\label{sd}
\begin{figure}
\begin{center}
\epsscale{1.0}
\plottwo{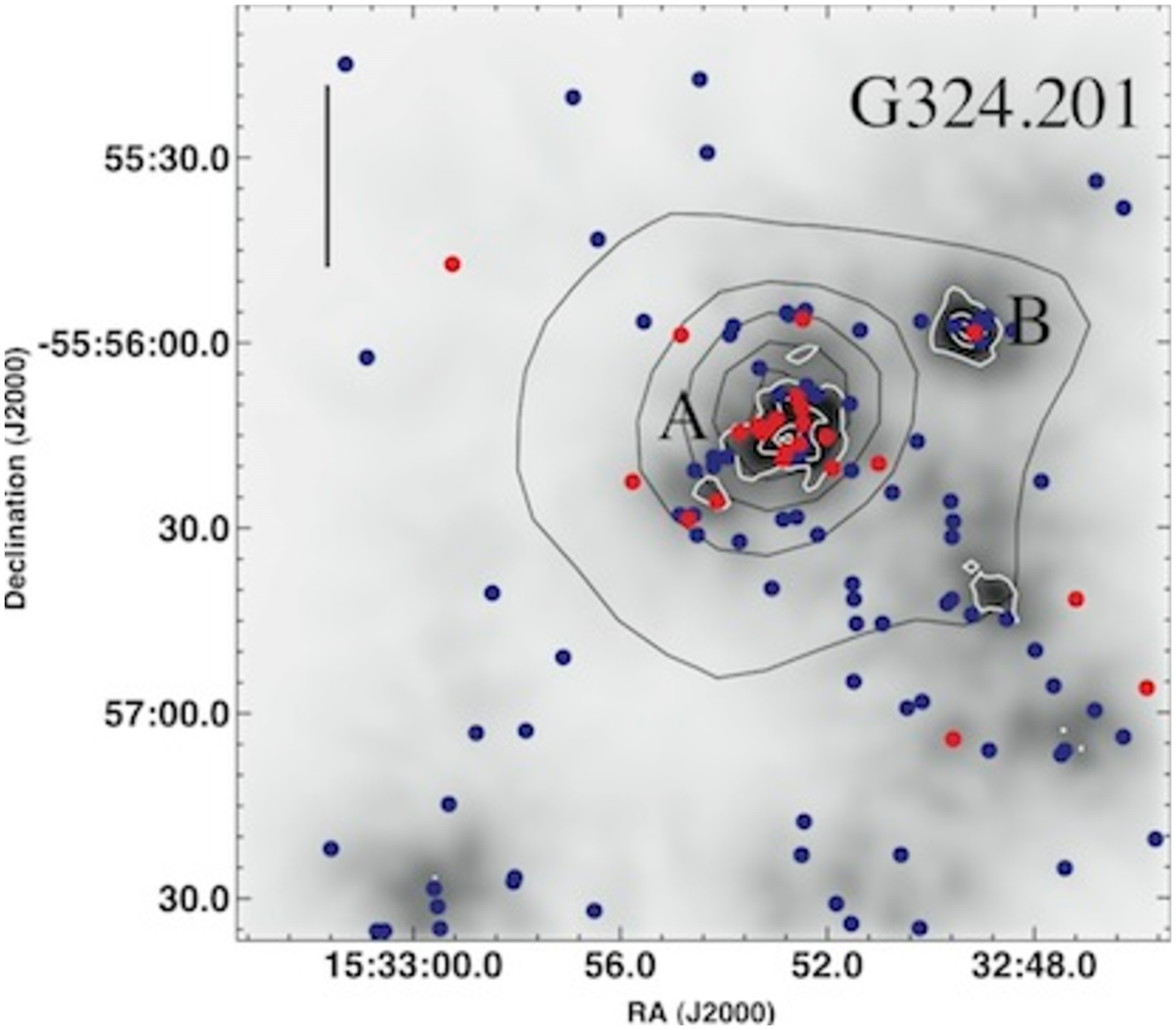}{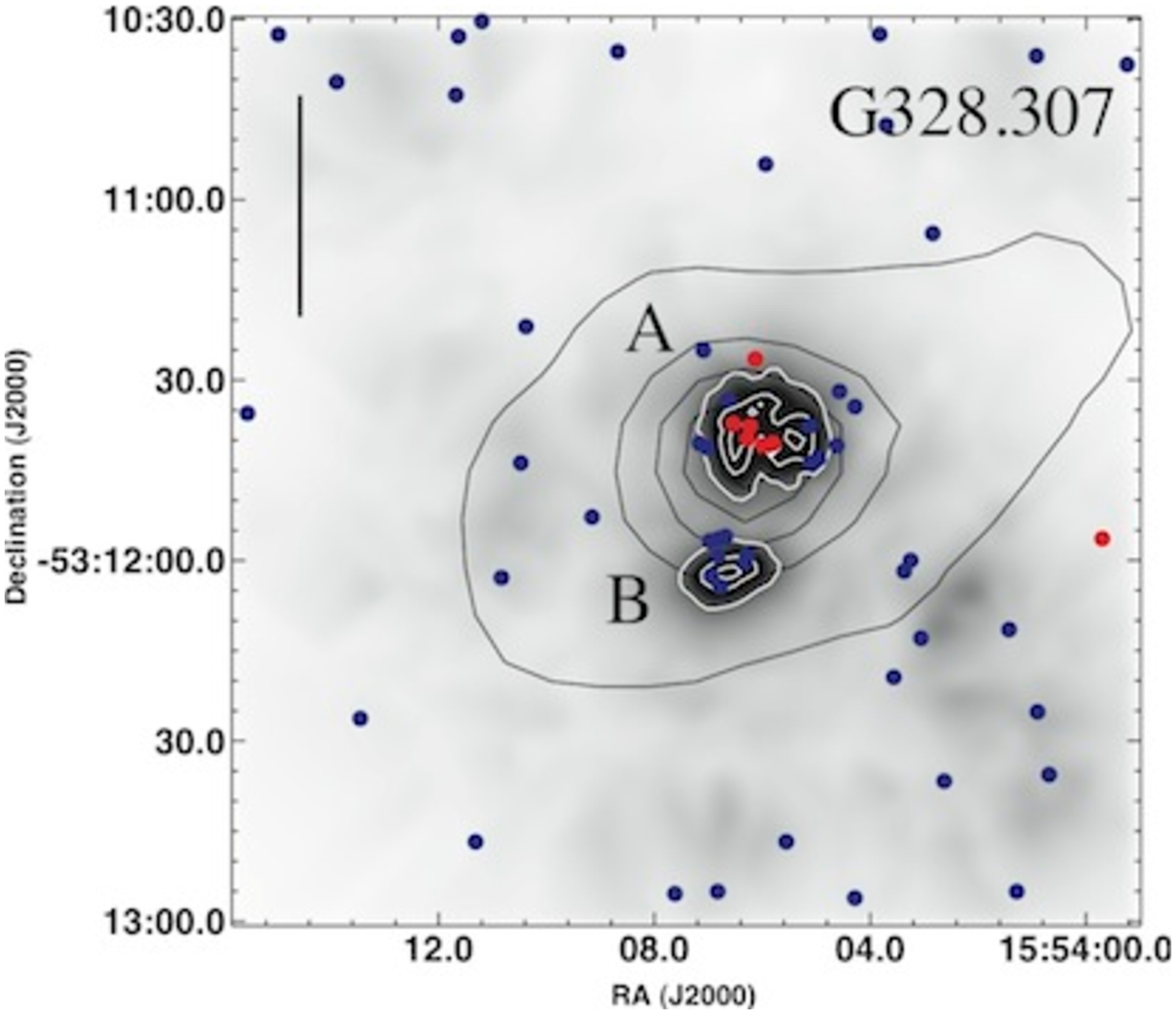}
\plottwo{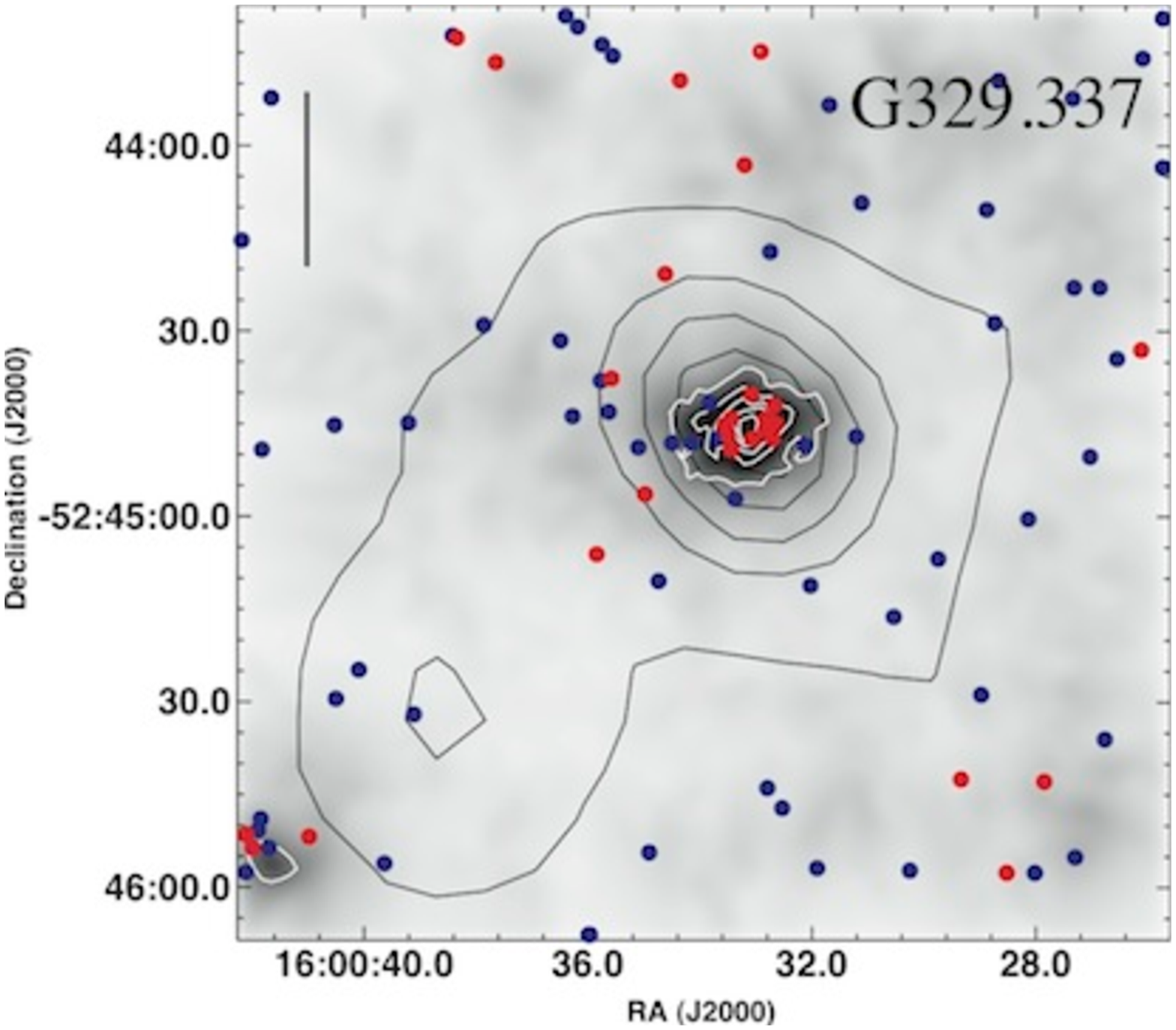}{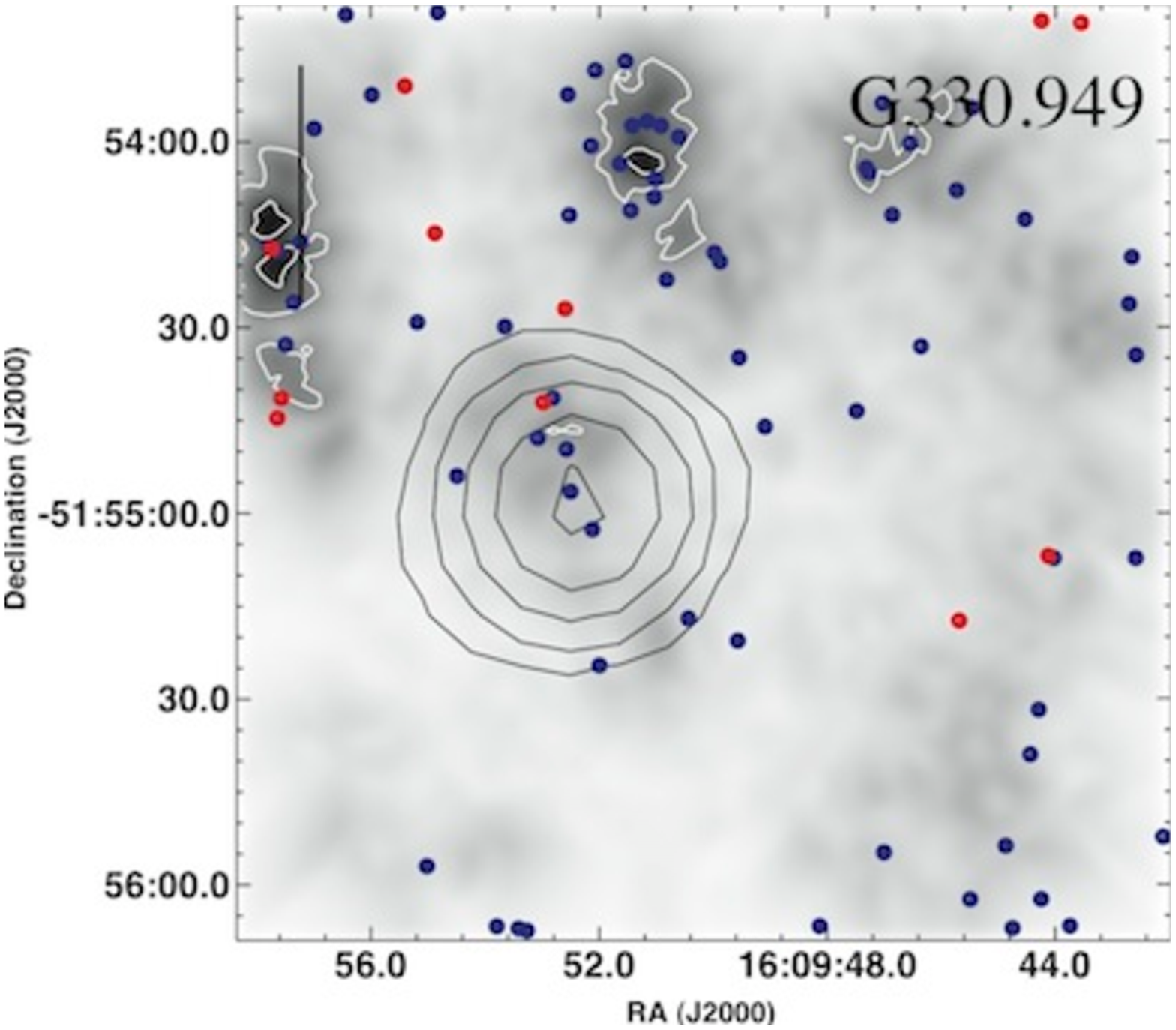}
\caption{Surface density maps of NIR-excess sources.  White contours begin at 3$\sigma$ over the median surface density (22, 11, 13 and 12~stars per pc$^{2}$ for G324.201, G328.307, G329.337 and G330.949 respectively) and increase by 3$\sigma$.  Massive and intermediate-mass stars with NIR-excess are marked with red and blue dots, respectively. Black contour levels correspond to 10, 30, 50, 70 and 90 \% of the peak dust emission detected with SIMBA at 1.2~mm \citep{gar09}. The straight line indicates a length of 1~pc.\label{surface_density}}
\end{center}
\end{figure}

Having identified the sources with NIR-excess, we can study their spatial distribution. This was done by dividing the mosaics using a regular grid and calculating the surface density of stars at each point of the grid:
\be
\sigma=\frac{N}{\pi r_{N}^2},
\ee
where $r_{N}$ is the distance to the $N=10$ nearest neighbor (NN). Figure~\ref{surface_density} shows the surface density maps of NIR-excess sources for all the regions. The maps were created using a 3 arcsecond grid and smoothed applying a convolution with a 3.0 arcsecond Gaussian kernel. The lower contour in the maps correspond to 3$\sigma$ over the mean value. The peak surface densities are 120, 40, 75 and 30 stars per pc$^{2}$ for G324.201, G328.307, G329.337 and G330.949, respectively.  

We use the surface density maps to find groups of NIR-excess sources. We define a group as a concentration of 10 or more NIR-excess sources with a surface density of stars higher than 3$\sigma$ over the mean. Using this definition, we found five groups: two towards G324.201 and G328.307 (labeled A and B) and one towards G329.337. Spurious group detections near the edges of the field in each region were ignored due to large photometric errors. Table~\ref{clusters} gives group parameters. Columns 2 and 3 list the coordinates corresponding to the peak surface density. Figure~\ref{surface_density} shows also the spatial distribution of \emph{massive} and \emph{intermediate-mass} sources with NIR-excess (in red and blue respectively). Groups G324.201A, G328.307A and G329.337 contain a large number of massive NIR-excess sources while G324.201B and G328.307B are mainly composed by intermediate and low-mass stars. The peak position of groups G324.201A, G328.307A and G329.337 is coincident, within the errors, with the peak position of the massive dense cores detected at 1.2~mm in Paper I. Groups G324.201B and G328.307B are found projected towards the massive and dense cores and are less prominent than their A companions. In what follows, we consider groups A and B as substructures of a single cluster, centered at the peak position of group A. The total number of cluster members, estimated by counting the NIR-excess sources inside a radius of 1.5~pc from the cluster center, is given in Table~\ref{IMnumber}. This radius corresponds roughly to the radius of the 10\% level of the peak 1.2~mm dust emission from the massive and dense cores (Paper I).

Since the NIR excess emission is likely to arise from disk-like structures, we assume 
that the NIR-excess sources are young and hereafter we will refer to them as YSOs. 
The spatial distribution of massive and intermediate-mass YSOs is discussed in \S~\ref{section_spatial01}. The non-detection of YSO groups in G330.949 is discussed in \S~\ref{section_g330}

\subsection{Spatial distribution of J-dropout sources}\label{section_spatial_Jdropout}
As mentioned in~\S~\ref{section_colmag}, the  J-dropout sources selected (brighter than a B2 star) likely correspond to embedded YSOs. If massive J-dropout sources correspond to either field stars or background AGB galaxies, we expect them to be uniformly distributed over the field. On the other hand, if they are associated with young clusters of NIR-excess sources discussed in \S~\ref{sd}, we expect them to have a similar spatial distribution.

\begin{figure}
\begin{center}
\epsscale{1.0}
\plottwo{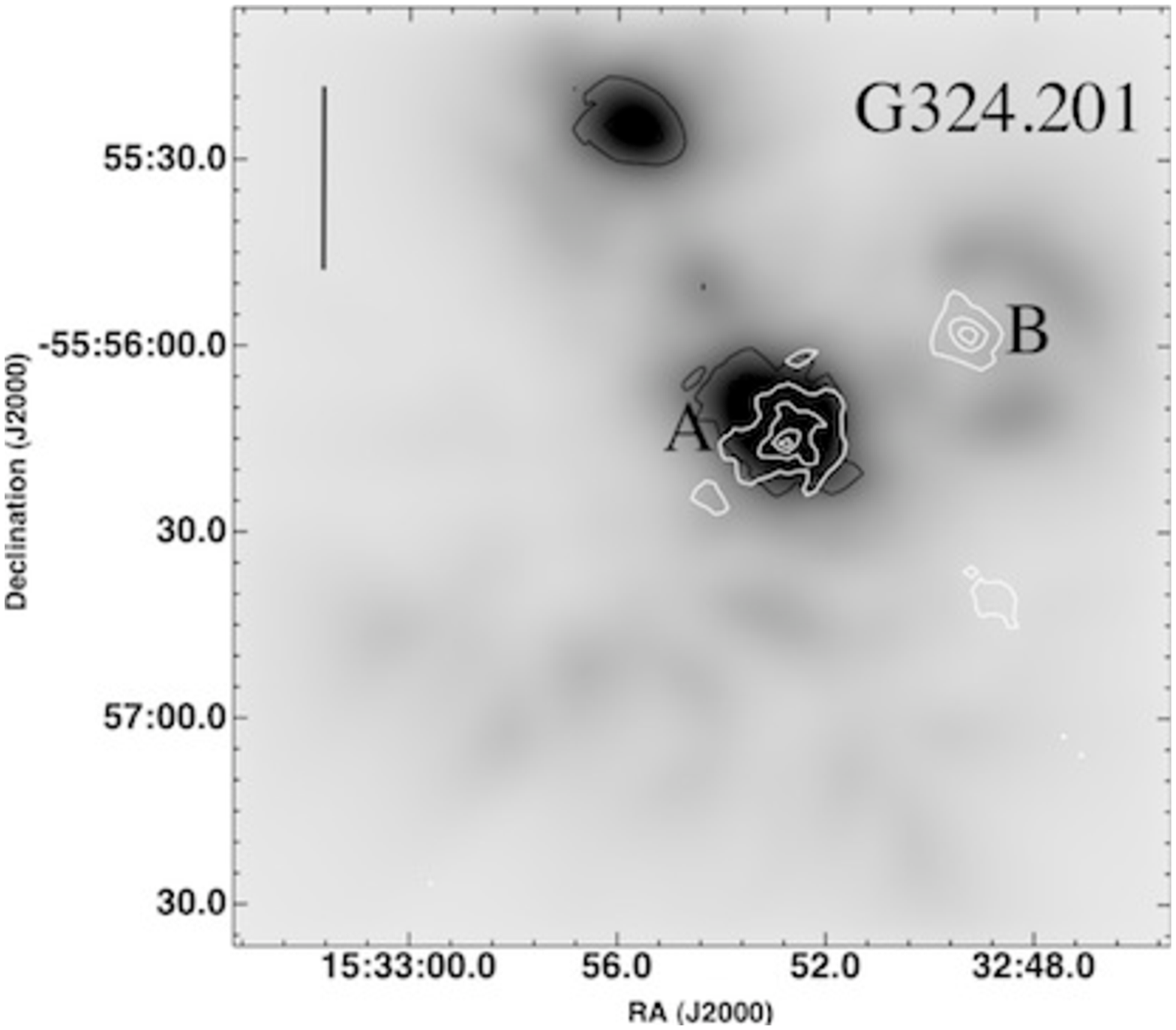}{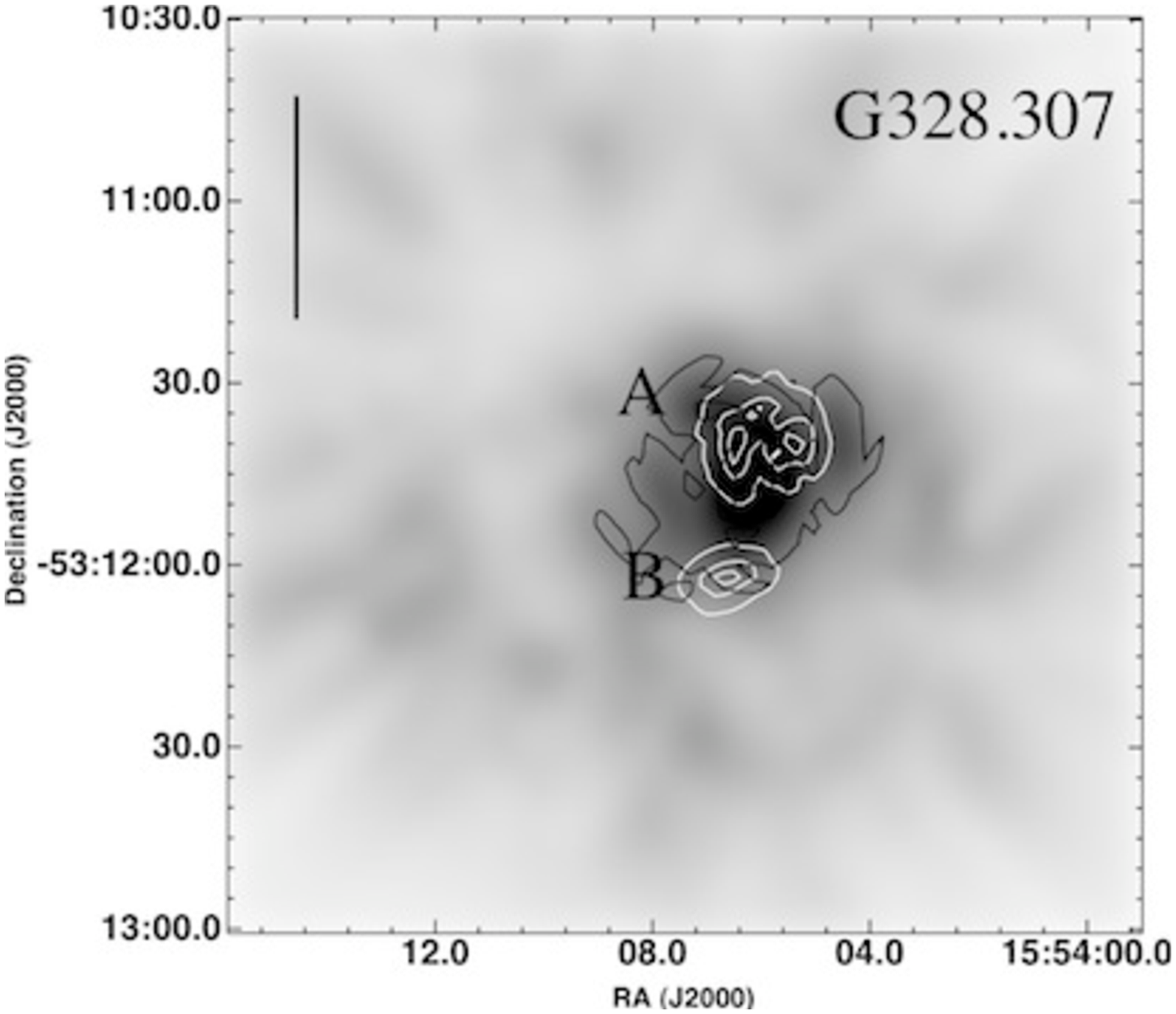}
\plottwo{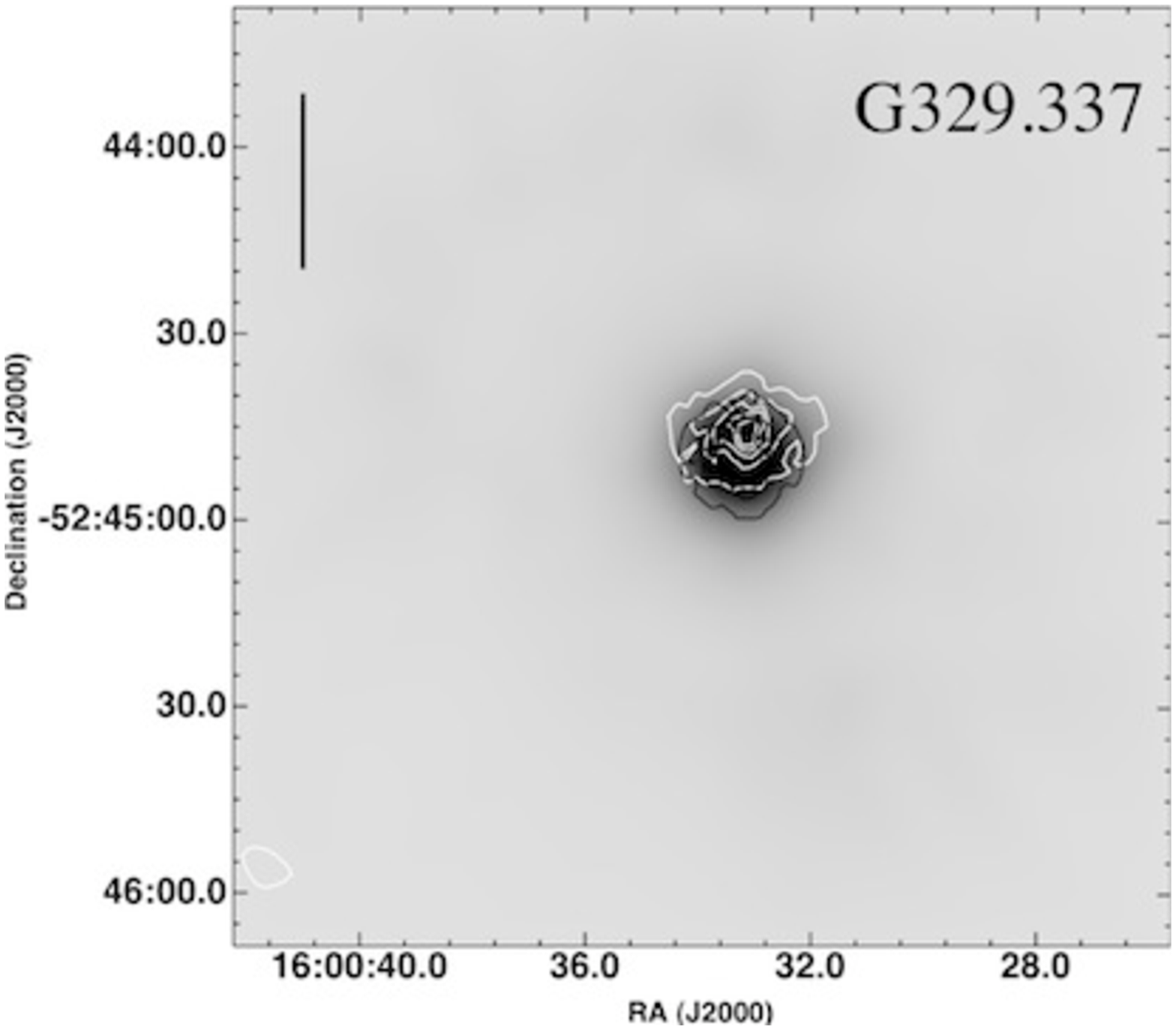}{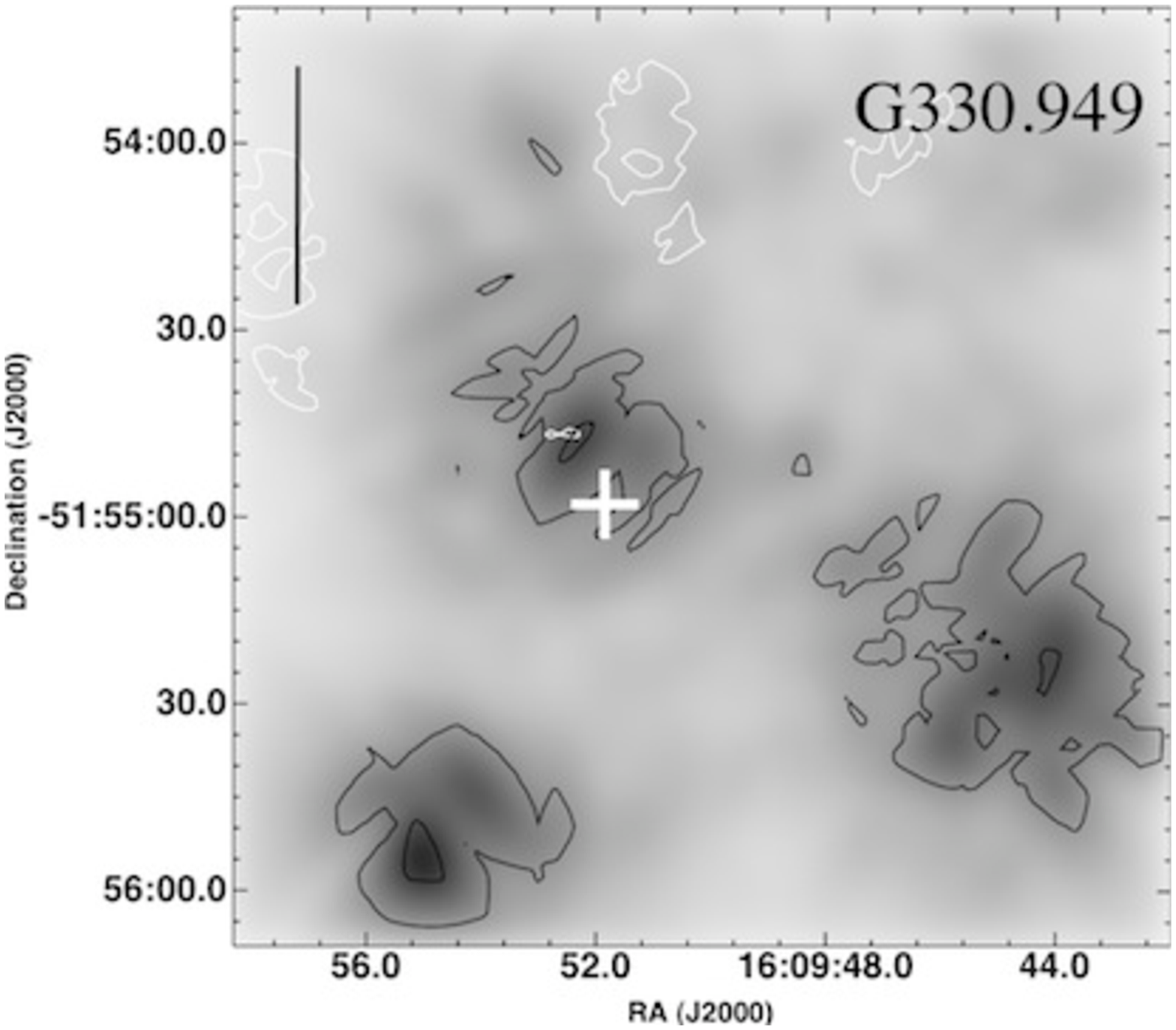}
\caption{Surface density maps of massive J-dropout sources with colors corresponding to spectral types earlier than a B2V star (shown as squares in Figure~\ref{color_magnitude}). Black contours begin at 3$\sigma$ over the mean surface density (9.3, 3.3, 15 and 2.25 stars per pc$^{2}$ for G324.201, G328.307, G329.337 and G330.949 respectively) and increase by 3$\sigma$. White contours correspond to the surface density of NIR-excess sources from Figure~\ref{surface_density}. The straight lines represent a distance of 1~pc. There is a clear spatial correspondence between massive J-dropout and NIR-excess sources in G324.201, G328.307 and G329.337. The white plus-sign in G330.949 indicates the position of the 1.2~mm emission peak from \citet{gar09}.\label{jdropout}}
\end{center}
\end{figure}

The surface density maps of massive J-dropout sources are shown in Figure~\ref{jdropout}. They were created using the same technique as for the NIR-excess surface density maps. Figure~\ref{jdropout} shows that far from being distributed homogeneously, the sources are concentrated at the location of clusters G324.201, G328.307 and G329.337, suggesting that those J-dropout sources are cluster members. These are probably very embedded sources and good protostar candidates.

\subsection{Extinction maps}
We compare the distribution of young stars with the distribution of dust.   We created extinction maps in each region using the mean value of the K-band extinction (A$_K$) of the nearest $N=10$ stars over a uniform 3.0 arcseconds grid. The A$_K$ values were estimated by derredening the sources in the color-color diagram to their main sequence colors. Sources with NIR-excess were derredened to the CTTS loci (see Figure~\ref{colcol}).  We convolved the resulting map with a 3.0 arcsecond Gaussian kernel. In order to eliminate the contribution of foreground stars to the extinction we used stars with A$_K > 1.0$ in G324.201 and G329.337 and with A$_K > 0.8$ in G328.307 and G330.949. This corresponds to the interstellar extinction at the distance to each region assuming an interstellar extinction of A$_K = 0.15$~magnitudes per kpc \citep{ind05}. It is important to note that due to the probable non detection of background stars located behind densest parts of the molecular clouds, the extinction values estimated represent a lower limit to the true extinction towards those regions. 

\begin{figure}
\begin{center}
\epsscale{1.0}
\plottwo{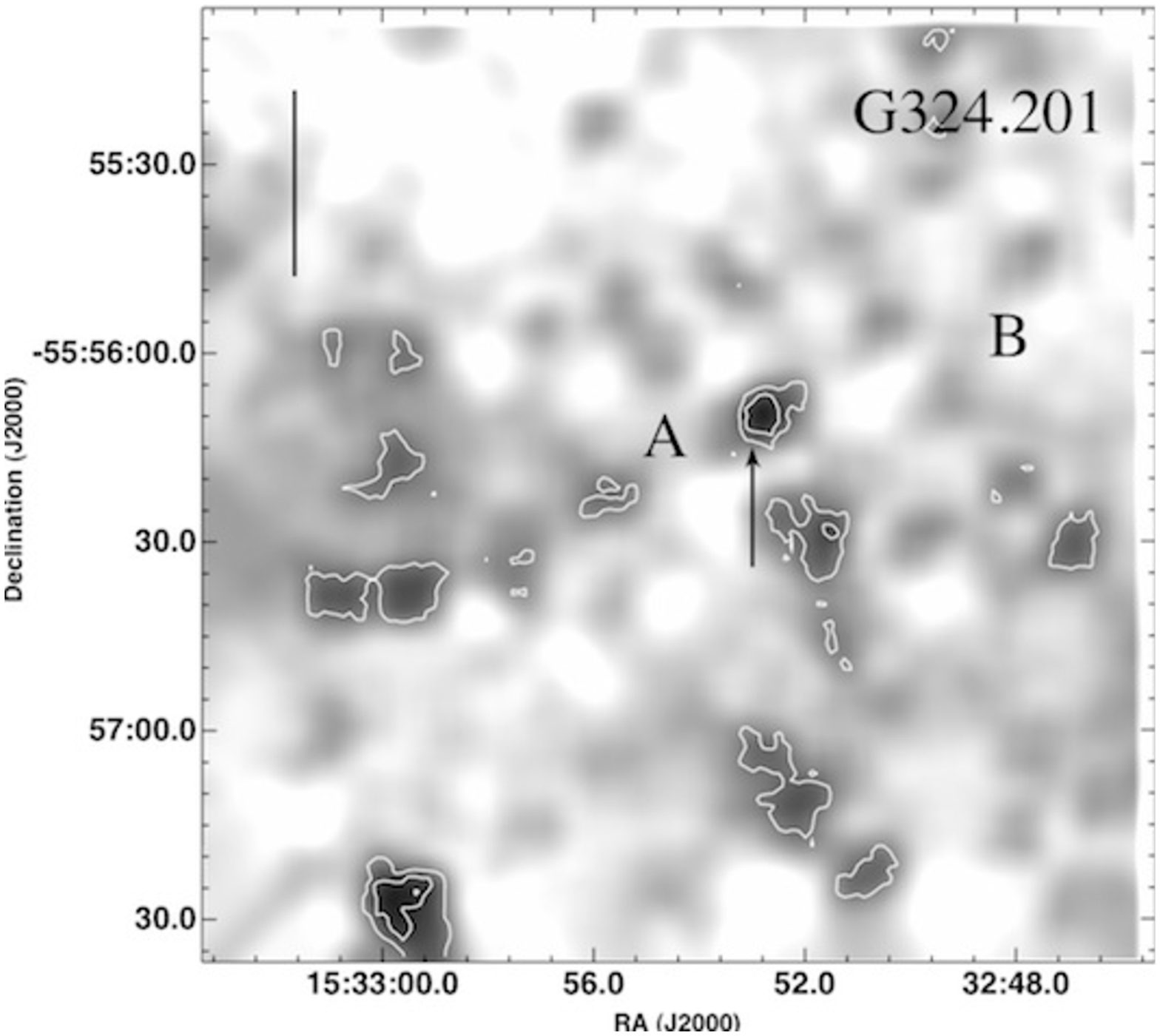}{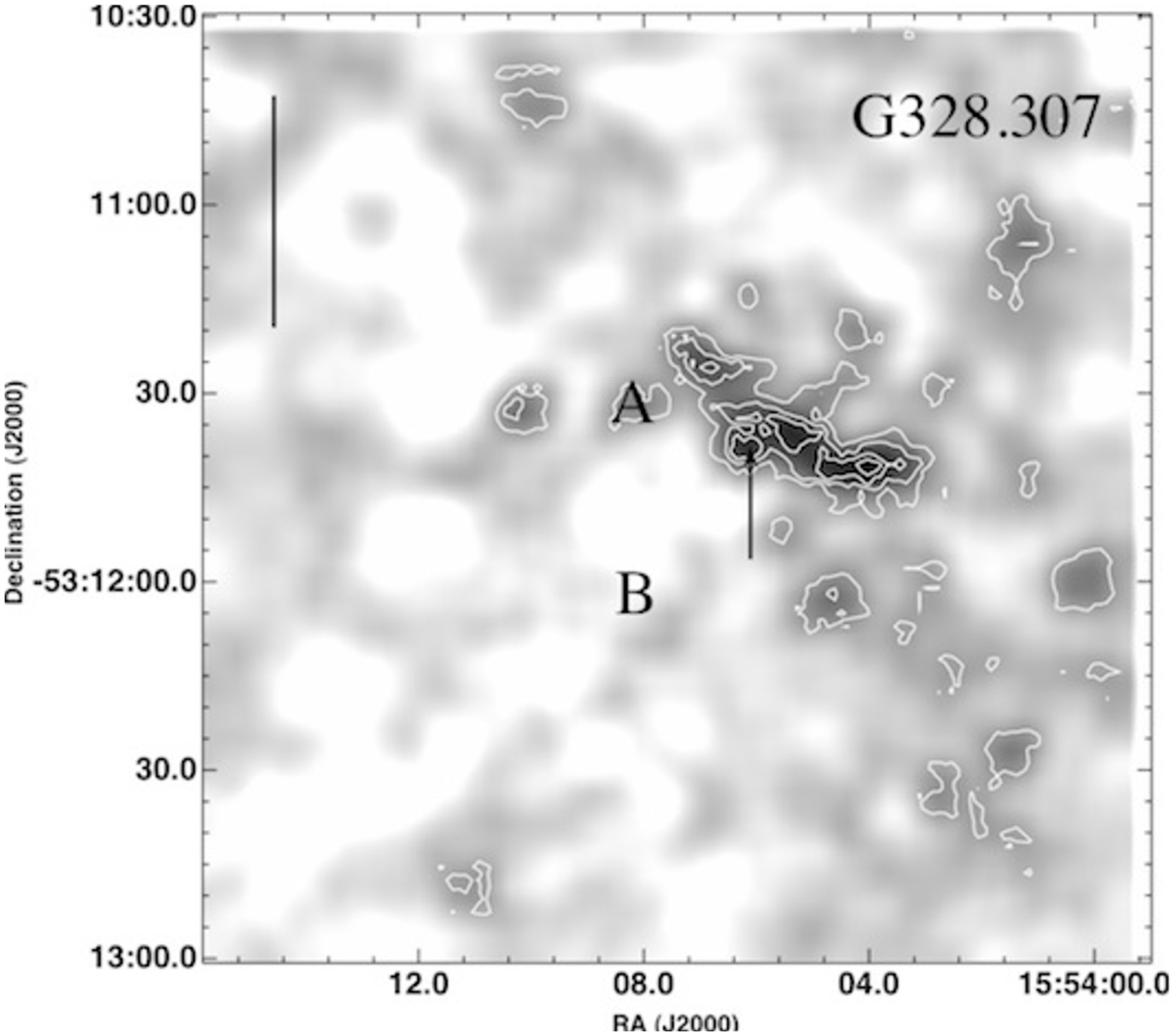}
\plottwo{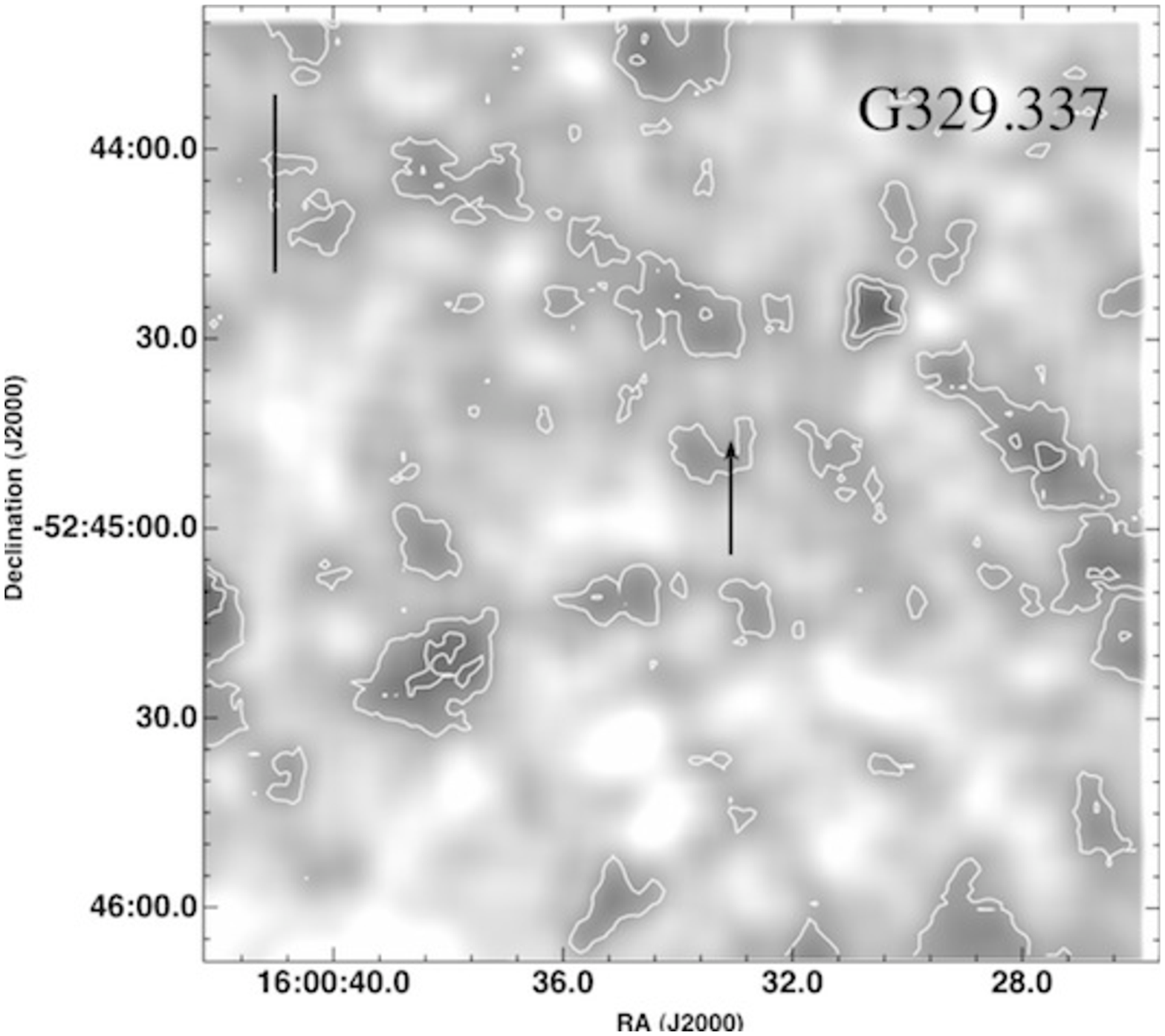}{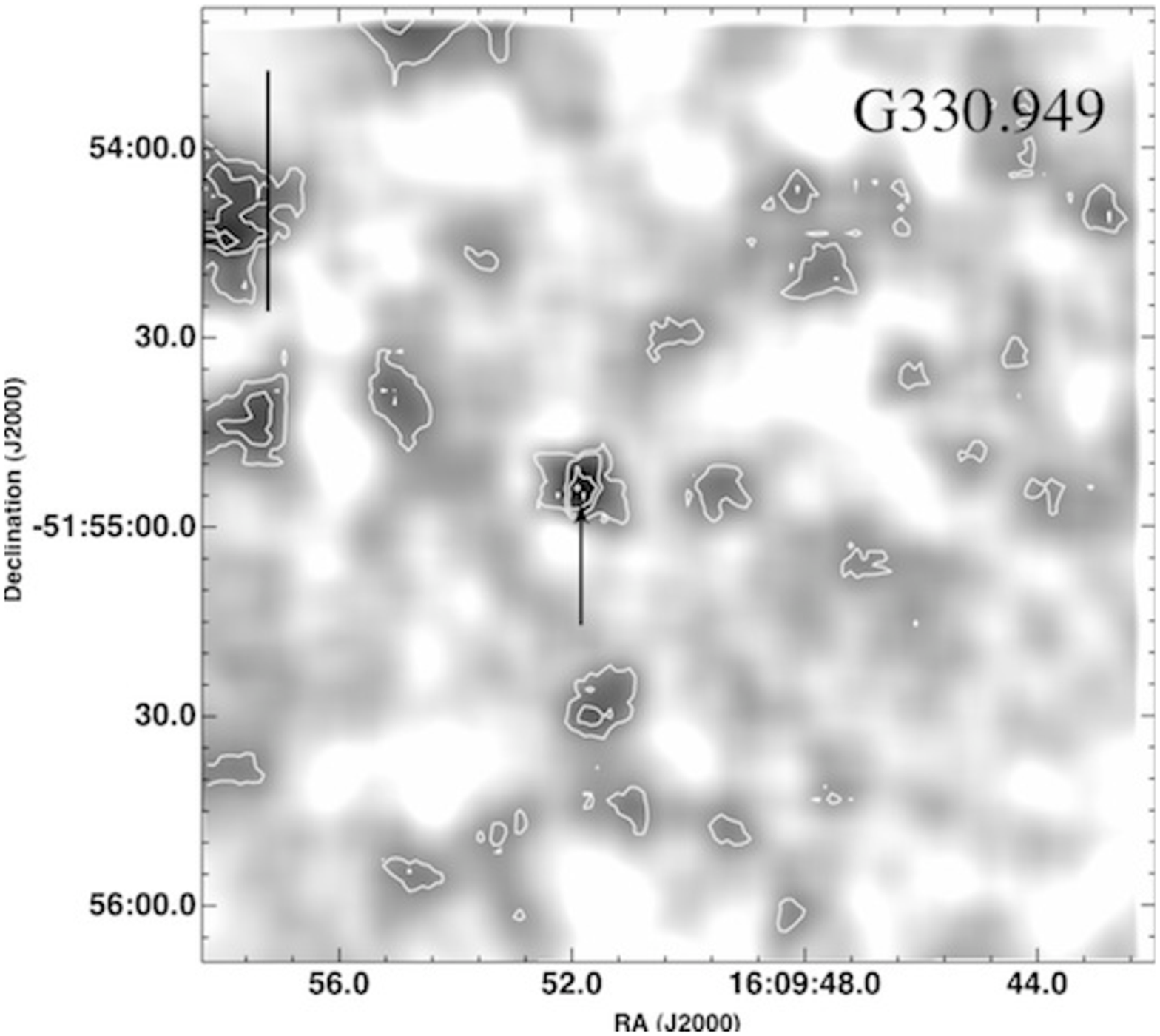}
\caption{K-band extinction maps for the observed regions. Contours begin at 1.8 A$_K$ and increase by 0.2.  The black lines represent a distance of 1~pc. The center of the clusters are indicated by the arrows. In G330.949, the arrow indicates the position of the 1.2~mm emission peak from \citet{gar09}.\label{extinction}}
\end{center}
\end{figure}

Figure~\ref{extinction} shows the K-band extinction map in each of the four regions. Groups G324.201A and G328.307A coincide with extinction peaks. The extinction peak in G330.949 is located at less than 1 arcsecond from the peak 1.2~mm emission (indicated by the arrow in Fig.~\ref{extinction}). The mean values of A$_K$ are close to 1.5 in all regions and the peak A$_K$ values are 2.2, 2.7, 2,2 and 2.6 for G324.201, G328.307, G329.337 and G330.949 respectively. These values are similar to the extinction in other sites of high mass star formation like the complex S254-S258 \citep{cha08a} and DR21/W75 \citep{kum07}. 

\section{Discussion}\label{section_discussion1}

\subsection{Radial distribution of NIR-excess sources inside the massive and dense cores}\label{section_spatial01}
\begin{figure}
\begin{center}
\epsscale{0.6}
\plotone{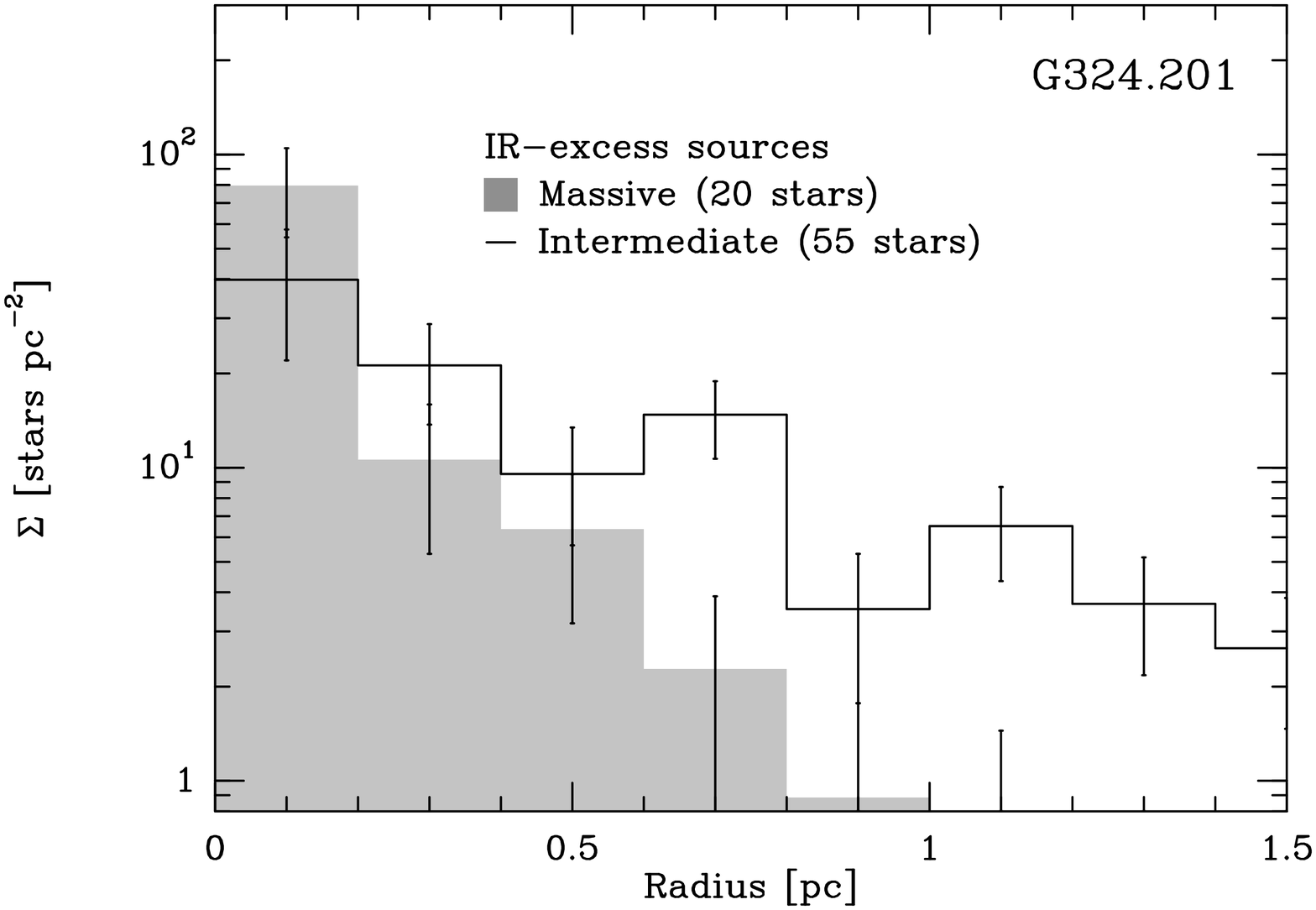}
\plotone{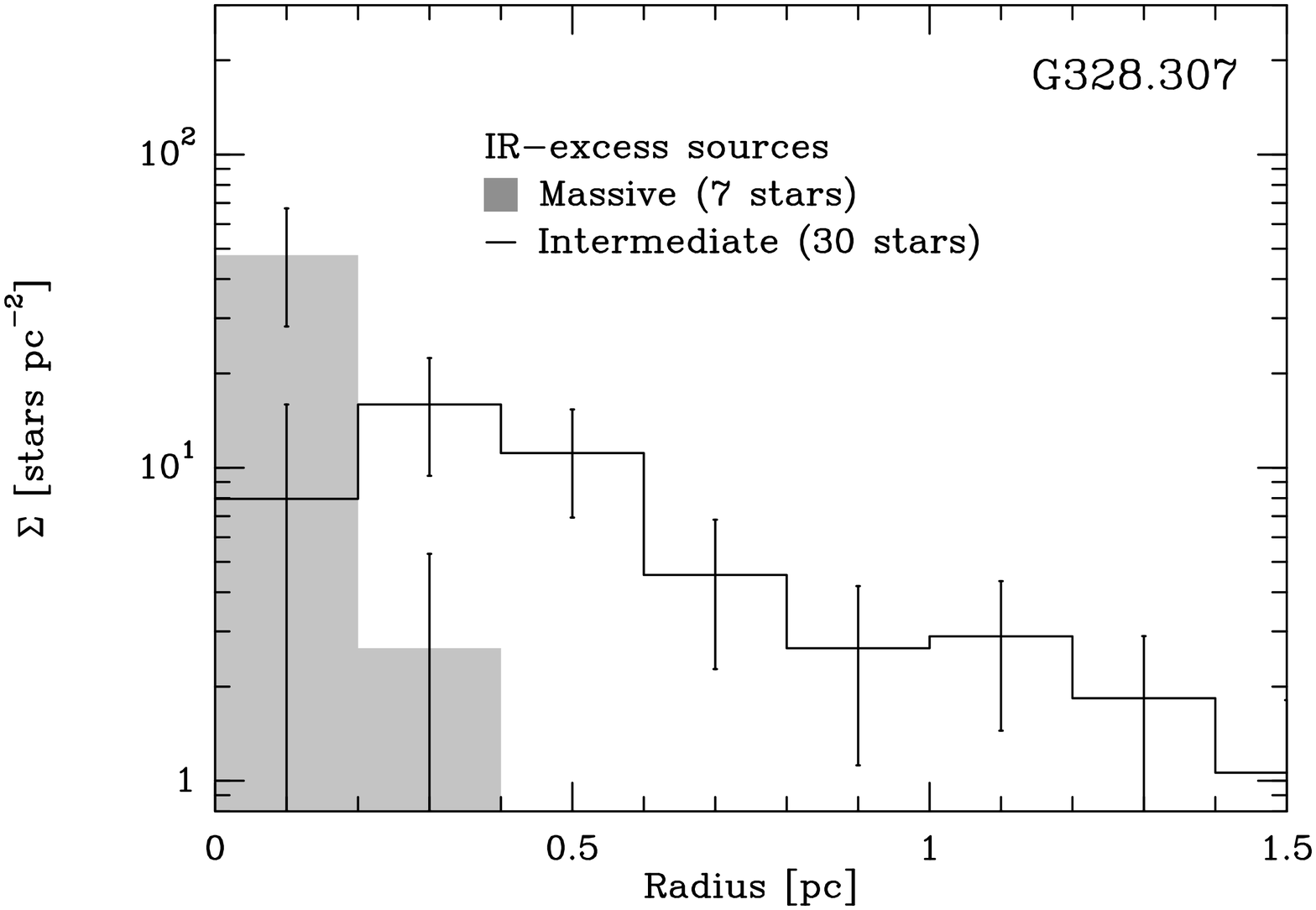}
\plotone{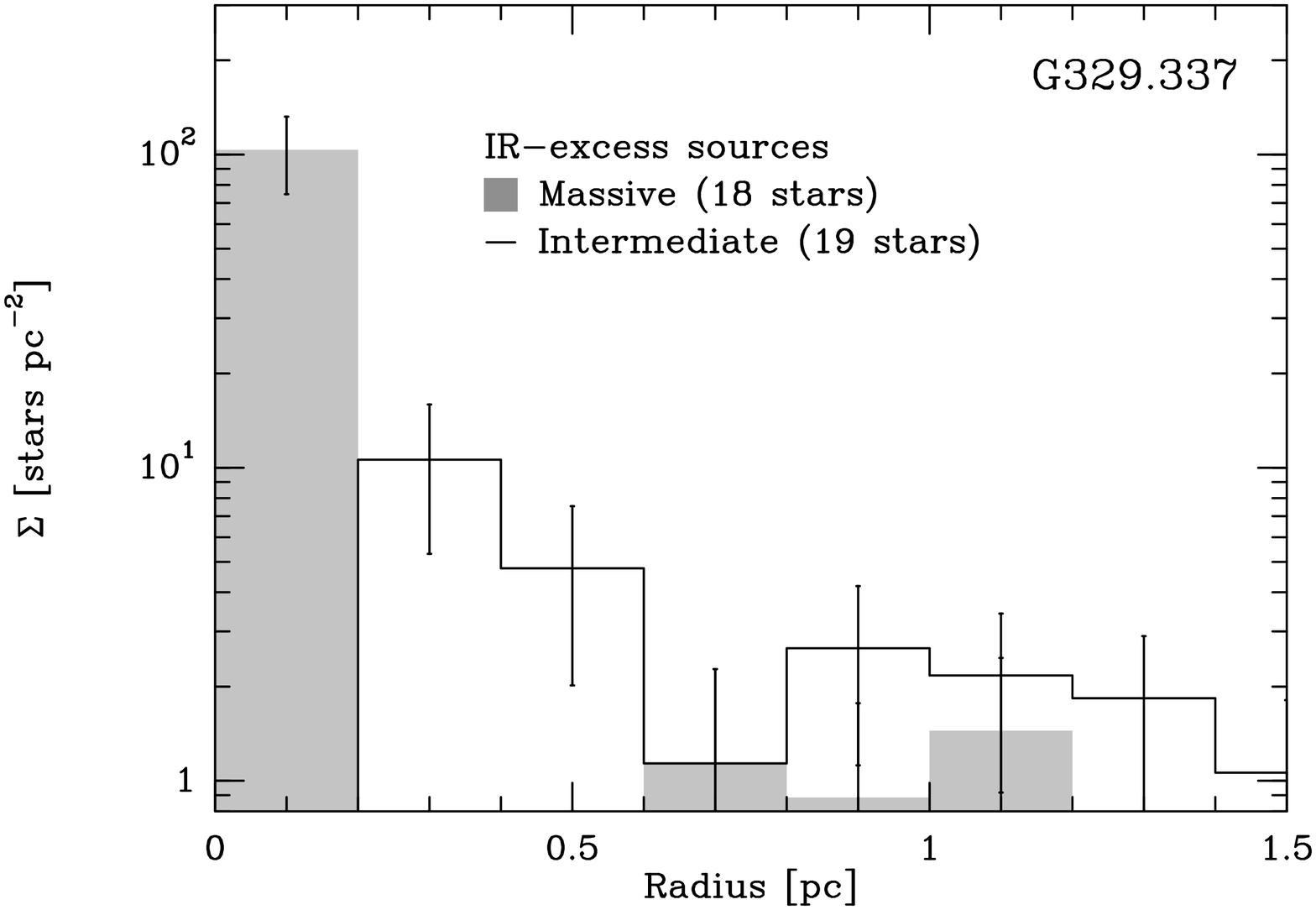}
\caption{Radial surface density of NIR-excess massive (gray) and intermediate-mass (line) sources in clusters G324.201, G328.307 and G329.337.\label{spatial_distribution}}
\end{center}
\end{figure}

In \S~\ref{redal} and \ref{section_colmag} we used the color-color diagram to identify NIR-excess sources and the color-magnitude diagram to divide stars into two groups depending on their estimated spectral type: massive and intermediate-mass stars. In this section, we investigate the spatial distribution of the NIR-excess sources from both groups within the massive and dense cores. We define the radial surface density of embedded sources as:
\be
\Sigma_i=\frac{N_i}{\pi(r^2_i - r^2_{i-1})},
\ee
where $N_i$ is the number of NIR-excess sources located between a radius of $r_{i-1}$ and $r_i$ from the center of the cluster. The step in radius used is 0.2~pc, about a third of the radius of the massive dense cores. Figure~\ref{spatial_distribution} shows $\Sigma_i$ histograms for \emph{massive} and \emph{intermediate-mass} YSO. The error bars in Figure~\ref{spatial_distribution} correspond to the square root of the number of YSO in each bin. The histograms show a clear difference in the radial distribution of massive and intermediate-mass YSO. In clusters G328.307 and G329.337, massive YSO are highly concentrated within the central region whereas intermediate-mass YSO are more spread. The mean distance to the cluster center for massive YSO in G328.307, $\langle r_{\mathrm{M}}\rangle$, is 0.09~pc and for G329.337 is 0.15~pc, while for intermediate-mass YSO the mean distances $\langle r_{\mathrm{I}}\rangle$ are 0.60 and 0.87~pc, respectively. 

Cluster G324.201 shows a concentration of both massive and intermediate-mass YSO in the core. To quantify how different are both distributions, we fitted the surface density $\Sigma$ with a radial power-law of the form:
\be
\Sigma = \left(\frac{r_0}{r_i}\right)^{-\alpha},
\ee
where $r_0$ is a constant. The values of $\alpha$ in G324.201 for massive and intermediate-mass YSO are 1.79 $\pm$ 0.06 and 0.77 $\pm$ 0.09, respectively, suggesting that massive IR-excess sources are more concentrated than intermediate-mass YSO.

\subsection{Dynamical effects and cluster ages}\label{section_relaxation_time}

In the previous section we concluded that there are signatures of mass segregation 
in the clusters G324.201, G328.307 and G329.337. In order to determine whether the 
segregation is primordial or dynamic (see \S~\ref{section_introduction}), we will 
compare the age of the cluster members with their relaxation time. One of the 
most used expressions to estimate the relaxation time of clusters is given 
by \citep{spi71,bin87};
\be
t_{relax} = \frac{6.5\times10^8}{\ln (0.4N)} \left(\frac{M}{10^5M_{\odot}}\right)^{1/2} \left(\frac{M_{\odot}}{m}\right) \left(\frac{r_{\textrm{\tiny{H}}}}{\textrm{pc}}\right)^{3/2}~\textrm{yr}\label{trelax},
\ee
where $N$ is the number of cluster members, $M$ is the total mass, $m$ is the mean mass of cluster members and $r_{\textrm{\tiny{H}}}$ is the half-mass radius (approximately 0.5~pc for all clusters). We calculated the number of cluster members by counting the YSOs inside a radius of 1.5~pc from the cluster center. The number of undetected cluster members was estimated using a synthetic cluster created with the algorithm from \citet{mue00}. We used a Trapezium IMF and age of 0.5-1.0~Myr as inputs. Based on the number of observed YSO, we used the synthetic cluster to estimate the percentage of cluster members with a magnitude fainter than the K-band completeness limit (see Table~\ref{completeness}). The corrected number of cluster members is then estimated by correcting the observed number of NIR-excess sources by the percentage of undetected members. Having the corrected number of cluster members, their mass is estimated assuming a mean mass of $m=0.5$ \msun~per star. Using equation (\ref{trelax}), we found relaxation times $t_{relax}$ between 5 and 8~Myr for cluster members in the studied dense cores. 

\begin{figure}
\begin{center}
\epsscale{0.6}
\plotone{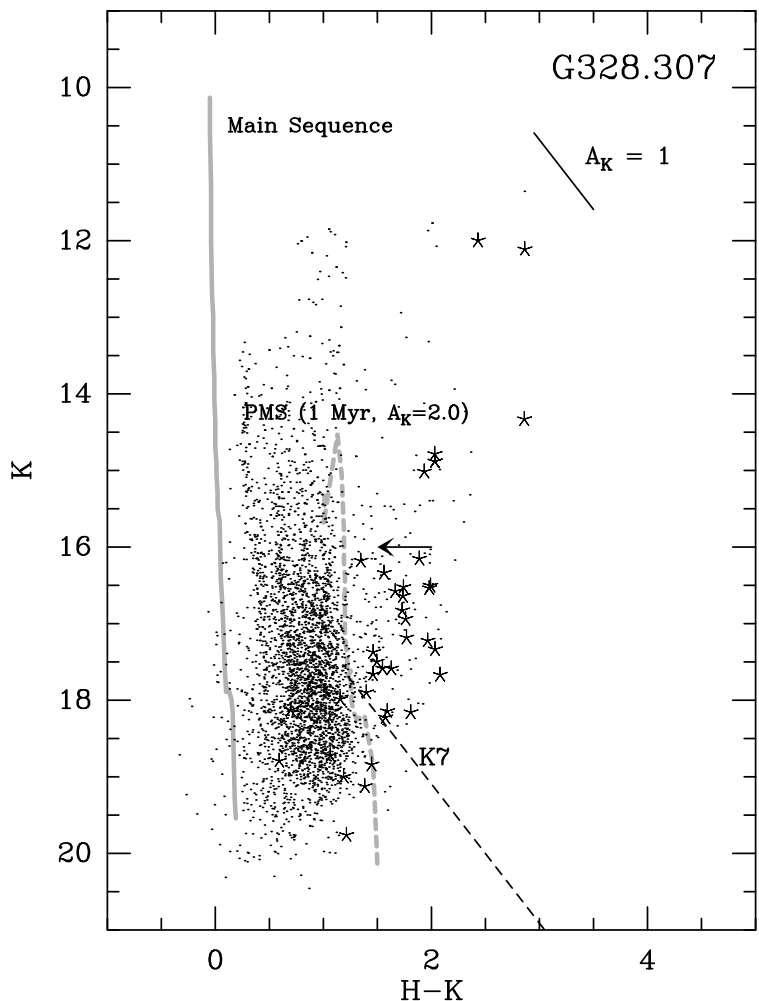}
\caption{Color-magnitude diagram for G328.307. Stars correspond to NIR-excess sources within 1~pc from the center of the cluster G328.307. The dot-dashed line indicates the 1~Myr pre-main sequence track from \citet{sie00}. We added an extinction of A$_K=2.0$ to the PMS model to resemble the physical conditions of the cluster. The dashed line corresponds to the reddening vector for a K7 PMS type star.  The arrow corresponds to an excess of 0.5 mag in the $H-K$ color, which is the mean excess of cluster members in G328.307. \label{colmagC}}
\end{center}
\end{figure}

In order to estimate the age of the clusters, we can use for instance the fact that they are still associated with their parental molecular cloud, implying that they are younger than 5~Myr \citep{lei89}. A more accurate estimate can be obtained using theoretical isochrones in the color-magnitude diagram. We consider NIR-excess sources within 1~pc from the center of the clusters and compare their position in the color-magnitude diagram with the 1~Myr pre-main sequence (PMS) isochrone from \citet{sie00}. To reproduce the cluster environment we added an extinction of $A_K=2.0$ to the isochrone. Figure~\ref{colmagC} illustrates this method for G328.307, showing that cluster 
members have ages of less or of the order of 1~Myr. Using the same method, the estimated ages of cluster members for G324.201 and G329.337 are also of the order of 1~Myr. This suggests that the mentioned stellar clusters are not relaxed yet and that equipartition of energy is not the origin of the observed mass segregation.

Since the dense cores are very young and mainly gaseous, with only a small percentage ($\sim$10\%) of the total mass in the form of stars (see  Table~\ref{IMnumber}), dynamical friction (also called gravitational drag) produced by the gaseous background onto the stars may play an important role in the orbital evolution of stars. Dynamical friction is defined as momentum loss by a massive perturber (in this case a star) due to the interaction with its own gravitationally induced gaseous wake. It was first studied by \citet{cha43} for a background medium composed only by stars, but it can be extended to the evolution of a massive perturber in a gaseous background \citep{ost99,san01,kim07}. In this case, we find that for a system of density $\rm \rho$ and Mach number $\cal M$, a massive perturber of mass $\rm M_{\rm star}$ and velocity  $\rm V_{\rm star}$ feels a gravitational drag given by \citep{ost99}:
\be
F_{\rm DF} = -4\pi\rho\left(\frac{G M_{\rm star}}{V_{\rm star}}\right)^{2}\times f\left({\cal M} \right)\label{fdf}
\ee
where $\rm f\left({\cal M} \right)$ is a dimensionless funtion than depends, in addition to the Mach number, on the maximum and minimum impact parameters \citep{ost99}. This formulae has been numerically tested, finding surprinsingly good agreement with numerical results when the maximum impact parameter equals to the diameter of the perturber's orbit \citep{kim07}.

Gravitational drag on a gaseous background has been applied to study the evolution and migration of stars in dense gaseous star forming clouds using numerical smooth particle hydrodynamics simulations \citep[][in \S~5.2]{esc03,esc04}. The simulations have initial conditions similar to the ones found in the dense cores studied: the cloud density profile is described by an isothermal sphere $(\rho(r) \propto r^{-1.8})$, with perturber having masses of 1~\% of the total gas mass ($\rm M_{star} \sim 20$~\msun~$ \approx 0.01$ M$_{gas}$). The characteristic timescale found in these simulations for migration of the stars by a factor of~$\sim$2-3 is given by \citep[][eq. 18]{esc04}: 
\be
{\rm \tau} = \left(\frac{\rm M_{gas}}{\rm 10 \, M_{\sun}}\right)^{-1/2} \, \left(\frac{\rm R}{\rm 0.01 \, pc}\right)^{3/2} \, 3.6 \times 10^{4} \, {\rm yr}, \label{eq6}
\ee
where M$_{\rm gas}$ is the total mass of the gas and R is the initial distance of the stars to the center of the cloud.

Using equation (6) for a typical cloud mass of $\rm 4 \times 10^{3}$~\msun~and initial distance of 0.4~pc, the migration timescale is $\rm \tau = 4.5 \times 10^{5}$~yr. This is comparable to the estimated ages of the mentioned clusters ($\sim$ 1~Myr). 
Furthermore, the dependence of the gravitational drag force on the perturber's mass (equation 5), gives migration timescales proportional to the inverse of the perturber's mass. 
This implies that intermediate mass stars have migration timescales typically a 
factor $\sim$10 longer than those of high mass stars and hence longer than 
the estimated age of the clusters ($\sim 10^6$~yr). Therefore, gravitational drag 
is considerably less effective for intermediate mass stars compared with the 
most massive ones. This suggest that in the mentioned stellar clusters only the 
massive stars were able to efficiently migrate towards the center due to 
gravitational drag, being a possible dynamical explanation for the origin of 
the observed mass segregation. 

\subsection{Sample incompleteness and clusters IMF}
The analysis performed in \S~\ref{section_spatial01} shows evidence of mass segregation in three massive and dense cores. In a recent paper, \citet{asc09} studied the effects of binning and data incompleteness on the most common diagnostics used to find mass segregation. They found that incompleteness due to crowding in the center of non-segregated clusters will mimic the observed evidence of mass segregation. This is because low-mass stars located near massive stars are blended from detection by the higher luminosity of their massive companion and remain undetected, producing a fictitious lack of low-mass stars in the center of the cluster.

Since the clusters studied in this paper have different properties than the 
synthetic clusters from \citet{asc09}, we can't use their results in order to 
correct the incompleteness of our data. Thus, we investigated how the different 
sources of incompleteness -- dust extinction, number of cluster members and 
crowding -- may affect our results. 
The effects of extinction are difficult to evaluate since extinction is unique 
for every star. The estimated number of cluster members (see~\S\ref{sd}) corresponds to a lower limit 
since we are not taking into account possible members without NIR-excess. 
\citet{cha08a} showed that the percentage of non-excess members in young clusters 
range roughly between 20-80\% for clusters with ages between 0.5-2~Myr. Since the 
clusters studied here are probably younger than 1~Myr, we expect that the 
percentage of non-excess members is roughly 50\% and therefore we are 
identifying approximately half of the cluster members.

\citet{asc09} showed that crowding will have an important effect near the center 
of the cluster. The identification of two or more stars as separate entities near 
the central region will be more difficult due to the small distance between cluster 
members. Two stars will be identified as separate entities if their projected 
distance is about two times the angular resolution. For our data, this translates 
roughly in a separation of 0.02~pc at the distances of the clusters 
(see~\S~\ref{photometry}). Assuming, for simplicity, that the stars near the 
cluster center are distributed homogeneously, then we will be able to identify 
separate entities if the density of stars is less than $2.5\times10^{3}$ stars per 
pc$^{3}$. This value is higher than the density of massive plus intermediate-mass 
stars in the central parsec of the dense cores ($\sim$ 150 stars per pc$^3$) even if we correct by the number of sources with no IR-excess, indicating that crowding is not an important source of incompleteness.

A possible way to test the evidence of mass segregation in our data is to calculate, for a given IMF, the number of intermediate-mass YSOs expected from the high-mass YSOs detected and compare it with the observed number. If both numbers agree, the data is complete and mass segregation is real. We define the IMF by:
\be
\frac{dN}{d(\log m)}= Am^{\Gamma},
\ee
where \emph{N} is the number of stars, \emph{m} is the mass, $\Gamma$ is the IMF slope and \emph{A} is a constant. From the number of high-mass NIR-excess sources, we estimate the number of intermediate-mass stars using two different IMF slopes: Salpeter \citep[$\Gamma=-1.35$,][]{sca86} and Arches \citep[$\Gamma=-0.9$,][]{sto05}. The mass range used for the high-mass stars is 30 to 8 \msun~and for the intermediate-mass is 8 to 2.1 \msun. Montecarlo simulations were used to estimate the probability of drawing N$_{\mathrm{IM}}$ intermediate-mass stars given the number N$_{\mathrm{M}}$ of massive stars from a mother distribution function described by a power-law with exponent $\Gamma=-1.35$ and $-0.90$. Table~\ref{IMnumber2} shows that the distribution of mass in G324.201 is better described by an Arches IMF, in G328.307 the distribution of mass is well described by both Salpeter and Arches IMF. This suggests that mass segregation in these clusters is real assuming a top-heavy IMF. 

The hypothesis of a top-heavy IMF is also suggested by the comparison of the gas mass with the mass in stars for different IMFs in dense cores. The mass of gas in dense cores was estimated in Paper I using two different methods: the CS emission line (\mcs) and the 1.2~mm dust emission (\mdust).  Both masses are in excellent agreement for all the dense cores. In addition, the total mass of the cores (gas plus stars,  \mvir) was also estimated, obtaining similar values to the mass of gas in cores. This means that the mass of stars is negligible compared to the mass of gas for most of the cores, which is also suggested by our estimation of the mass of stars (see Table~\ref{IMnumber}). If we use the number of cluster members with masses of more than 8~\msun~to estimate the total mass of stars for different IMF slopes, we find that for a Salpeter IMF, the mass of stars is about half of the mass of gas and non negligible (see Table~\ref{masses}). This contradiction suggest that there is a lack of low-mass stars in the studied clusters and supports the idea of a top-heavy IMF.

\subsection{Different main sequences in color-magnitude diagram}\label{section_multipleMS}
\begin{figure}
\begin{center}
\epsscale{0.6}
\plotone{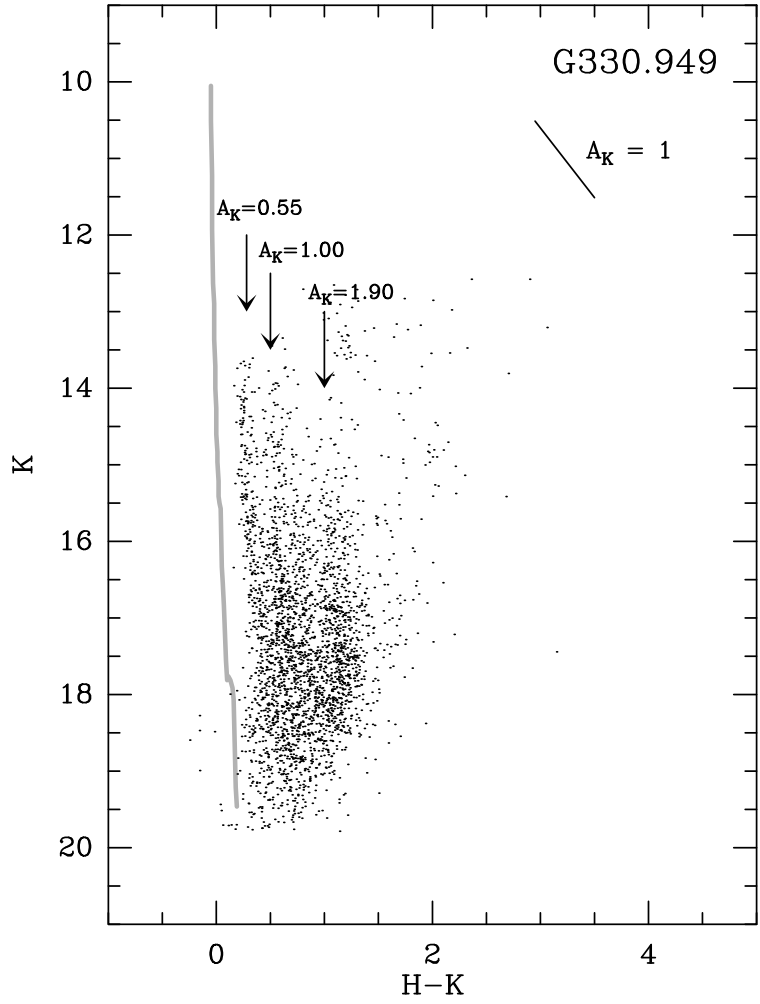}
\caption{Color-magnitude diagram for G330.949. We only included sources detected in J, H and K bands. Arrows indicate the extinction of the reddened main sequences at A$_K$ of 0.55, 1.0 and 1.9.\label{colmagB}}
\end{center}
\end{figure}

In this section we investigate the origin of the multiple main sequences seen in the color-magnitude diagrams of G328.307, G329.337 and G330.949 (see~Figure~\ref{color_magnitude}). All the studied sources are located near the tangent of the Norma spiral arm. along those lines of sight there are two other spiral arms at closer distances: Carina and Scutum-Crux. Molecular clouds that belong to those spiral arms are likely to cover the field of view adding layers of dust. They will increase the extinction of background stars which will appear as reddened main sequences in the color-magnitude diagram. Figure~\ref{colmagB} shows three main sequences at different extinction in the color-magnitude diagram of G330.949 (NIR-excess sources were plotted as points to distinguish the different main sequences). 

We correlate the column density derived from the extinction of the different main sequences in the color-magnitude diagrams with the column density derived from molecular material in the line of sight to the clusters. The \coa~survey of \citet{bro89} shows that the clusters have a \vlsr~of the order of -120 to -80~km~s$^{-1}$, but it also shows that there is molecular material with \vlsr~$> -80$~km~s$^{-1}$ along the line of sight to all the regions. In the case of G330.949, there are two molecular components; one with \vlsr~$\sim -65$~km~s$^{-1}$ and other with $\sim -45$~km~s$^{-1}$. We estimated the H$_2$ column density from the extinction in the K-band using the relation given by \citet{dic78}:
\be
\frac{N(H_2)}{A_V} = 1.25\times 10^{21}~cm^{-2}~mag^{-1}\label{dust2.1}
\ee
and the relation between visual and K-band extinction from \citet{car89}:
\be
\frac{A_K}{A_V} = 0.114.\label{dust2.2}
\ee
By combining equations~(\ref{dust2.1}) and~(\ref{dust2.2}) we obtain:
\be
\frac{N(H_2)}{A_K}=1.1\times10^{22}~cm^{-2}~mag^{-1}.\label{column_density}
\ee
We used the $A_K$ values of the main sequences in the color-magnitude diagram (see Figure~\ref{colmagB}) and equation~(\ref{column_density}) to estimate the H$_2$ column density. The values estimated are 0.58$\times 10^{22}$~cm$^{-2}$ for A$_K=0.55$ and 0.48$\times 10^{22}$~cm$^{-2}$ for A$_K=0.45$. On the other hand, H$_2$ column density values derived from the molecular component were taken directly from \citet{bro89}, being in agreement by a factor of 2 with the column densities derived from A$_K$. A comparison of the H$_2$ column density values derived from both methods is shown in Table~\ref{table_density_column} for G328.307, G329.337 and G330.949. We conclude that the reddened main sequences in the color-magnitude diagrams are probably due to the presence of layers of dust along the line of sight.

\subsection{The stellar population within the G330.949 molecular core}\label{section_g330}
\begin{figure}
\begin{center}
\epsscale{0.8}
\plotone{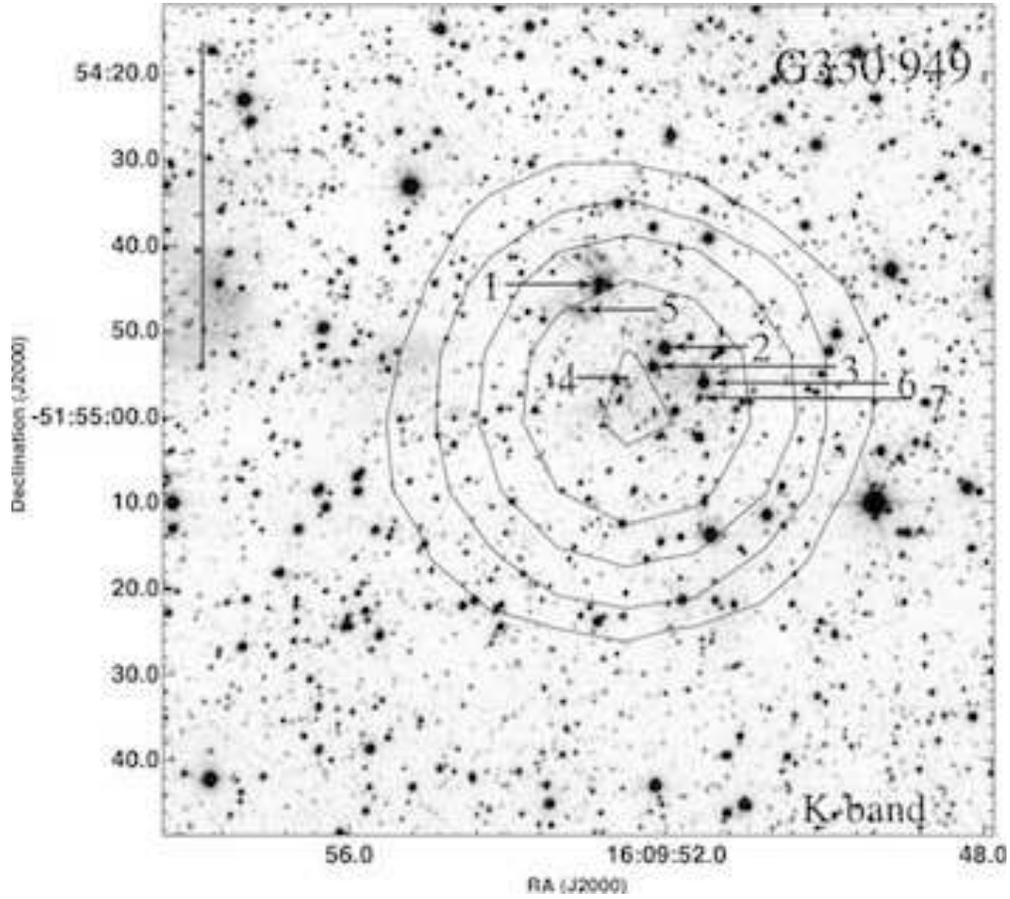}
\caption{K-band image of G330.949. Seven sources of interest are shown in the figure (see explanation in~\S~\ref{section_g330}). Contours correspond to 10, 30, 50, 70 and 90\% of the 1.2~mm dust emission (from Figure~\ref{surface_density}).  The black line represents 1 pc.\label{g330}}
\end{center}
\end{figure}

As discussed in \S~\ref{sd}, the G330.949 molecular core is the only core in which 
we did not detect a group of NIR-excess sources, even though this region is the most luminous at far-IR wavelengths (see Table~\ref{sources}). 
To investigate the main sources of luminosity in this region
we examined individual objects within the contours of the 1.2~mm emission. 
We found seven objects, labeled in Figure~\ref{g330}, with characteristics commonly associated to star formation: 
NIR-excess, extended emission and high extinction. Objects 2 and 3 are the most 
extinted stars in the region as can be seen in the color-color diagram 
(Fig.~\ref{colcol}). Objects 1 and 4 are J-dropouts and object 5 has NIR-excess.
In addition, objects 1 and 5 are associated with extended emission. 
Most of these objects are concentrated near the peak position of the 1.2~mm 
emission suggesting that star formation is taking place near the center of 
G330.949. The position of objects 1, 2 and 3 in the (K vs. H-K) color-magnitude 
diagram suggests that they correspond to highly extincted O6 stars. If so, 
their total luminosity is $\sim7\times10^5$ \lsun, indicating that they 
are responsible for most of the luminosity of the region. Objects 1, 2-3 and 6-7 correspond to objects 1, 2 and 3 from \citet{ost97}. 
They identified these objects as highly reddened sources in agreement with our results.

In addition, radio continuum observations with high ($\sim1$\arcsec) angular 
resolution towards IRAS 16060-5146 showed the presence of a compact HII 
region with an irregular morphology and a peak flux density of 280 mJy at 
8.6 GHz \citep{wal98}. It lies at the peak position of the dust core and 
is not associated with any of the NIR sources in our images, suggesting 
that the exciting source of the \hii~region is extremely embedded and 
undetectable at NIR wavelengths.

\section{Conclusions}\label{section_conclusions1}
We used NIR data from the VLT-ISAAC instrument to identify the young population of 
stars in four regions of massive star formation: G324.201, G328.307, G329.337 
and G330.949. These regions are located in the Norma spiral arm at distances 
between 5.4 and 7.3~kpc from the Sun. We used color criteria to identify sources 
with NIR-excess and found 236, 120, 186 and 228 NIR-excess sources in G324.201, 
G328.307, G329.337 and G330.949 respectively.

We identified clusters of YSOs in surface density maps of NIR-excess sources.
We found new clusters towards G324.201, G328.307 and G329.337. The spatial distribution 
of sources detected in H and K, but not in the J-band (J-dropout), 
was investigated. We concluded that J-dropout sources with colors of massive stars 
are likely to correspond to cluster members.

We analyzed the spatial distribution of massive and intermediate-mass cluster 
members and found evidence of mass segregation in each of the clusters. 
The effects of data incompleteness on the detected mass segregation 
were assessed and concluded that they are not important.

The age of the clusters was compared with their relaxation time and migration 
timescales due to gravitational drag, in order to determine if mass segregation is primordial or dynamical. We found that the migration timescales for massive stars is comparable with the age of the cluster members. This evidence suggest that gravitational drag from the gaseous component may be a possible cause for mass segregation in clusters G324.201, G328.307 and G329.337.

Several main sequences in the color-magnitude diagram of regions G328.307, G329.337 and G330.949 were identified. We computed the H$_2$ column density derived from \coa~for molecular components located in Scutum-Crux arm and compared those values with the column density derived from A$_K$ for the different main sequences. The H$_2$ column densities derived from both methods are in reasonable agreement, indicating that the reddening is mainly due to molecular clouds located along the line of sight.



\acknowledgments
This work was supported by CONICYT-CHILE through projects FONDAP 15010003 and BASAL PFB-06.

Facilities: \facility{VLT(ISAAC)}.





\clearpage

\clearpage
\begin{deluxetable}{ccccccccc}
\tabletypesize{\footnotesize}
\tablecaption{Source list \label{sources}}
\tablewidth{0pt}
\tablehead{
\colhead{Source} & 
\colhead{IRAS id.} & 
\colhead{lon} &
\colhead{lat} &
\colhead{$\alpha$(J2000)} &
\colhead{$\delta$(J2000)} &
\colhead{d} &
\colhead{\vlsr\tablenotemark{a}} &
\colhead{L$_{\mathrm{IRAS}}$} \\
& & \colhead{(deg)} & \colhead{(deg)} & \colhead{hh mm ss} & \colhead{dd mm ss} & \colhead{(kpc)} & \colhead{(km s$^{-1}$)} & \colhead{($L_{\odot}$)}
}
\startdata
G324.201 & 15290-5546 & 324.201 & 0.119  & 15 32 52 & -55 56 19 & 6.9\tablenotemark{b} & -88.0 & $7.8\times 10^5$\\
G328.307 & 15502-5302 & 328.307 & 0.432  & 15 54 05 & -53 11 37 & 5.6\tablenotemark{c} & -91.7 & $1.1\times 10^6$\\
G329.337 & 15567-5236 & 329.337 & 0.147  & 16 00 32 & -52 44 52 & 7.3\tablenotemark{b} & -107.1 & $1.3\times 10^6$\\
G330.949 & 16060-5146 & 330.949 & -0.174 & 16 09 52 & -51 54 38 & 5.4\tablenotemark{c} & -89.7 & $1.0\times 10^6$\\
\enddata
\tablenotetext{a}{From \citet{bro96}.}
\tablenotetext{b}{Tangent distance to Norma arm.}
\tablenotetext{c}{Near side distance to Norma arm \citep{bro00}.}
\end{deluxetable}

\clearpage
\begin{deluxetable}{crrr}
\tablecaption{Integration Times \label{time}}
\tablewidth{0pt}
\tablehead{
\colhead{Source} & 
\colhead{J$_{\textrm{s}}$\tablenotemark{a}} &
\colhead{H\tablenotemark{a}} &
\colhead{K$_{\textrm{s}}$\tablenotemark{a}} 
}
\startdata
G324.201 & 1200  & 2400 & 3600 \\
G328.307 & 1200  & 2400 & 3600 \\
G329.337 &  300  &  300 &  300 \\
G330.949 &  300  &  300 &  300 \\
\enddata
\tablenotetext{a}{Time in seconds.}
\end{deluxetable}

\clearpage
\begin{deluxetable}{cccccccc}
\tablecaption{Number of detected stars\tablenotemark{a}\label{stars}}
\tablewidth{0pt}
\tablehead{
\colhead{Source} & 
\colhead{J$_{\textrm{s}}$} &
\colhead{H} &
\colhead{K$_{\textrm{s}}$} &
\colhead{J-dropout\tablenotemark{b}} &
\colhead{J$_{\textrm{s}}$HK$_{\textrm{s}}$} &
\colhead{M\tablenotemark{c}} &
\colhead{IM\tablenotemark{d}}
}
\startdata
G324.201 & 4809 & 5721 & 8347 & 1081 & 4057 &  99 & 1449 \\
G328.307 & 4103 & 6036 & 7057 & 1349 & 3779 &  64 & 1002 \\
G329.337 & 4431 & 6357 & 7927 & 1424 & 3876 &  82 & 1411 \\
G330.949 & 3516 & 5512 & 6887 & 1580 & 2894 &  49 &  773 \\
\enddata
\tablenotetext{a}{Detections with error $<$ 0.2 mag.}
\tablenotetext{b}{Sources detected in H and K bands but no detected in J band.}
\tablenotetext{c}{Massive: stars with spectral type earlier than B2.}
\tablenotetext{d}{Intermediate-mass: stars with spectral type between B2 and A5.}
\end{deluxetable}

\clearpage
\begin{deluxetable}{cccc}
\tablecaption{90\% Completeness\label{completeness}}
\tablewidth{0pt}
\tablehead{
\colhead{Source} & 
\colhead{J$_{\textrm{s}}$} &
\colhead{H} &
\colhead{K$_{\textrm{s}}$} \\
& \colhead{(mag)}& \colhead{(mag)}& \colhead{(mag)}
}
\startdata
G324.201 & 20.2 & 17.5 & 17.2 \\
G328.307 & 20.7 & 18.5 & 17.5 \\
G329.337 & 19.5 & 17.5 & 16.5 \\
G330.949 & 19.3 & 16.8 & 16.7 \\
\enddata
\end{deluxetable}

\clearpage
\begin{deluxetable}{lccrrr}
\tablecaption{Group properties\label{clusters}}
\tablewidth{0pt}
\tablehead{
\colhead{Name} & 
\colhead{$\alpha$(J2000)} &
\colhead{$\delta$(J2000)} &
\colhead{N$_{\mathrm{IR}}$\tablenotemark{a}} &
\colhead{N$_{\mathrm{M}}$\tablenotemark{b}} &
\colhead{N$_{\mathrm{IM}}$\tablenotemark{c}} \\
& \colhead{hh mm ss} & \colhead{dd mm ss} &&&
}
\startdata
G324.201A & 15 32 52.8 & -55 56 15.7  & 28 & 12 & 9\\
G324.201B & 15 32 49.3 & -55 55 58.2  & 15 &   1 & 5\\
G328.307A & 15 54 06.5 & -53 11 39.9  & 20 &   7 & 6\\
G328.307B & 15 54 06.6 & -53 12 02.0  & 10 &   0 & 5\\
G329.337   & 16 00 33.1 & -52 44 45.6  & 26 & 12 & 5\\
\enddata
\tablenotetext{a}{Number of NIR-excess sources.}
\tablenotetext{b}{Number of massive NIR-excess sources.}
\tablenotetext{c}{Number of intermediate-mass NIR-excess sources.}
\end{deluxetable}

\clearpage
\begin{deluxetable}{lcccc}
\tablecaption{Cluster  parameters\label{IMnumber}}
\tablewidth{0pt}
\tablehead{
\colhead{Cluster} & 
\colhead{N$_{\mathrm{IR}}$\tablenotemark{a}} &
\colhead{N$_{\mathrm{cor}}$\tablenotemark{b}} &
\colhead{M$_{\mathrm{stars}}$} &
\colhead{\mdust\tablenotemark{c}} \\
&&& \colhead{[\msun]} & \colhead{[\msun]} 
}
\startdata
G324.201 & 130 &   890 & 445 & 5.0$\times 10^3$ \\
G328.307 &   60 &   345 & 173 & 5.6$\times 10^3$ \\
G329.337 &   62 & 1240 & 620 & 5.8$\times 10^3$ \\
\enddata
\tablenotetext{a}{Number of NIR-excess sources inside a radius of 1.5~pc from the cluster center.}
\tablenotetext{b}{Corrected number of members (see~\S~\ref{section_relaxation_time})}
\tablenotetext{c}{From Paper I}
\end{deluxetable}

\clearpage
\begin{deluxetable}{lcccccc}
\tablecaption{Expected number of intermediate-mass members in dense cores\label{IMnumber2}}
\tablewidth{0pt}
\tablehead{
\colhead{Cluster} & 
\colhead{N$_{\mathrm{M}}$\tablenotemark{a}} &
\colhead{N$_{\mathrm{IM}}$\tablenotemark{b}} &
\colhead{NIM$_{\mathrm{Salpeter}}$\tablenotemark{c}} &
\colhead{p$_{-1.35}$\tablenotemark{d}} &
\colhead{NIM$_{\mathrm{Arches}}$\tablenotemark{e}} &
\colhead{p$_{-0.90}$\tablenotemark{f}} 
}
\startdata
G324.201 & 20 & 55 & 122  & 0.03  & 67 & 0.49\\
G328.307 &   7 &  30 &   43  & 0.43 & 23  & 0.51\\
G329.337 & 18 & 19 & 110  & 0.00  & 60 & 0.02\\
\enddata
\tablenotetext{a}{Number of massive IR-excess sources.}
\tablenotetext{b}{Number of intermediate-mass IR-excess sources.}
\tablenotetext{c}{Expected intermediate-mass stars for $\Gamma=-1.35$.}
\tablenotetext{d}{Probability of drawing N$_{\mathrm{IM}}$ given N$_{\mathrm{M}}$ for a $\Gamma=-1.35$ IMF.}
\tablenotetext{e}{Expected intermediate-mass stars for $\Gamma=-0.90$.}
\tablenotetext{f}{Probability of drawing N$_{\mathrm{IM}}$ given N$_{\mathrm{M}}$ for a $\Gamma=-0.90$ IMF.}
\end{deluxetable}

\clearpage
\begin{deluxetable}{lcccccc}
\tablecaption{Mass of stars and gas in dense cores\label{masses}}
\tablewidth{0pt}
\tablehead{
\colhead{Cluster} & 
\colhead{N\tablenotemark{a}} &
\colhead{M$_{\mathrm{Salpeter}}$\tablenotemark{b}} &
\colhead{M$_{\mathrm{Arches}}$\tablenotemark{c}} &
\colhead{\mcs \tablenotemark{d} } & 
\colhead{\mdust \tablenotemark{d} } & 
\colhead{\mvir \tablenotemark{d} } \\
& & \colhead{[\msun]} & \colhead{[\msun]} & \colhead{[\msun]} & \colhead{[\msun]} & \colhead{[\msun]}
}
\startdata
G324.201 &   43 & 3000 & 1700 & 4.3$\times 10^3$ & 5.0$\times 10^3$ & 3.7$\times 10^3$\\
G328.307 &   43 & 3000 & 1700 & 4.6$\times 10^3$ & 5.6$\times 10^3$ & 6.8$\times 10^3$\\
G329.337 &   60 & 4100 & 2300 & 6.0$\times 10^3$ & 5.8$\times 10^3$ & 6.0$\times 10^3$\\
G330.949 &     -  &         - & -         & 1.0$\times 10^4$ & 1.1$\times 10^4$ & 7.9$\times 10^3$\\
\enddata
\tablenotetext{a}{Number of massive IR-excess plus J-dropout sources inside a 1.5~pc radius.}
\tablenotetext{b}{Mass of stars between 0.6 and 30~\msun~for $\Gamma=-1.35$.}
\tablenotetext{c}{Mass of stars between 0.6 and 30~\msun~for $\Gamma=-0.90$.}
\tablenotetext{d}{From Paper I.}
\end{deluxetable}

\clearpage
\begin{deluxetable}{ccccc}
\tablecaption{H$_2$ column density derived from \coa~and A$_K$.\label{table_density_column}}
\tablewidth{0pt}
\tablehead{
\colhead{Source} & 
\colhead{\coa\tablenotemark{a}} &
\colhead{$A_K$\tablenotemark{a}} &
\colhead{\coa\tablenotemark{b}} &
\colhead{$A_K$\tablenotemark{b}} \\
&  
\colhead{10$^{22}$[cm$^{-2}$]} & 
\colhead{10$^{22}$[cm$^{-2}$]} & 
\colhead{10$^{22}$[cm$^{-2}$]} & 
\colhead{10$^{22}$[cm$^{-2}$]} 
}
\startdata
G328.307 & 0.78 & 0.53 & 0.30 & 0.65 \\
G329.337 & 0.73 & 0.85 & 0.50 & 0.65 \\
G330.949 & 1.05 & 0.58 & 0.57 & 0.48 \\
\enddata
\tablenotetext{a}{Component with \vlsr~between -30 and -57~km~s$^{-1}$}
\tablenotetext{b}{Component with \vlsr~between -57 and -80~km~s$^{-1}$}
\end{deluxetable}


\begin{thebibliography}{}
\bibitem[Ascenso et al.(2007)]{asc07} Ascenso, J., Alves, J., Beletsky, Y. \& Lago, M., T., 2007, A\&A, 466, 137
\bibitem[Ascenso et al.(2009)]{asc09} Ascenso, J., Alves, J. \& Lago, M., T., 2009, A\&A, 495, 147
\bibitem[Balog et al.(2004)]{bal04} Balog, Z., Kenyon, S., Lada, E., Barsony, M., Vink\'o, J. \& G\'aspa\'r, A., 2004, AJ, 128, 2942 
\bibitem[Binney \& Tremaine(1987)]{bin87} Binney J. \& Tremaine, S., 1987, Galactic Dynamics, Princeton University Press
\bibitem[Blanc et al.(2008)]{bla08} Blanc G., et al., 2008, ApJ, 681, 1099
\bibitem[Bonnell et al.(2001)]{bon01} Bonnell, I., Bate, M., Clarke, J. \& Pringle, J., 2001, MNRAS, 323, 785
\bibitem[Bragg \& Kenyon(2005)]{bra05} Bragg, A., \& Kenyon, S., 2005, ApJ, 130, 134
\bibitem[Bronfman et al.(1996)]{bro96} Bronfman L., Nyman L. \& May J., 1996, A\&A, 115, 81.
\bibitem[Bronfman et al.(1988)]{bro88} Bronfman L., Cohen R. S., Alvarez H., May J.\& Thaddeus P., 1988, ApJ, 324, 248.
\bibitem[Bronfman et al.(1989)]{bro89} Bronfman L., Alvarez H., Cohen R. S.\& Thaddeus P., 1989, ApJS, 71, 481
\bibitem[Bronfman et al.(2000)]{bro00} Bronfman L., Casassus S., May J. \& Nyman L., 2000, A\&A, 358, 521.
\bibitem[Cardelli et al.(1989)]{car89} Cardelli, J., Clayton, C. \& Mathis, J., 1989, ApJ, 345, 245 
\bibitem[Chandrasekhar(1943)]{cha43} Chandrasekhar, S., 1943, ApJ, 97, 255 
\bibitem[Chavarr\'{\i}a et al.(2008)]{cha08a} Chavarr\'{\i}a, L., Allen, L., Hora, J., Brunt, C. \& Fazio, G., 2008, ApJ, 682, 445 
\bibitem[Chen et al.(2007)]{che07} Chen, L., de Grijs, R. \& Zhao, J., 2007, ApJ, 134, 1368 
\bibitem[Dickman(1978)]{dic78} Dickman, R., 1978, ApJS, 37, 407
\bibitem[Escala et al.(2003)]{esc03} Escala, A., Larson, R., Coppi, P. \& Mardones, D., 2003, in ASP Conference Series, 2002 International Astronomical Observatories in Chile workshop, ed. James M. De Buizer and Nicole S. van der Bliek, (San Francisco, CA: ASP), 287, 345 
\bibitem[Escala et al.(2004)]{esc04} Escala, A., Larson, R., Coppi, P., Paolo S. \& Mardones, D., 2004, ApJ, 607, 765
\bibitem[Kenyon et al.(1998)]{ken98} Kenyon, S., J., Lada, E., A. \& Barsony, M., 1998, ApJ, 115, 252.
\bibitem[Fa\'undez et al.(2004)]{fau04} Fa\'undez, S., Bronfman, L., Garay, G., Chini, R., Nyman, L. \& May, J., 2004, A\&A, 426, 97 
\bibitem[Garay et al.(1993)]{gar93} Garay, G., Rodr\'{\i}guez, L. F., Moran, J. M. \& Churchwell, E. 1993, ApJ, 418, 368.
\bibitem[Garay et al.(2007)]{gar07} Garay, G., Mardones, D., Brooks, K., Videla, L., Contreras, Y., 2007, ApJ, 666, 309.
\bibitem[Garay et al.(2009)]{gar09} Garay, G., Mardones, D., Bronfman, L., May, J. \& Chavarr\'{\i}a, L., 2009, submitted.
\bibitem[Gouliermis et al.(2004)]{gou04} Gouliermis, D., Keller, S., Kontizas, M., Kontizas, E. \& Bellas-Velidis, I., 2004, A\&A, 416, 137
\bibitem[Ho \& Haschick(1981)]{ho81} Ho, P. \& Haschick, A., 1981, ApJ, 248, 622 
\bibitem[Indebetouw et al.(2005)]{ind05} Indebetouw, R., et al., 2005, ApJ, 619, 931
\bibitem[Jiang et al.(2002)]{jia02} Jiang, Z., et al., 2002, ApJ, 577, 245
\bibitem[Kim \& Kim(2007)]{kim07} Kim, H. \& Kim, W-T., 2007, ApJ, 665, 432.
\bibitem[Koornneef(1983)]{koo83} Koornneef, J., 1983, A\&A, 128, 84.
\bibitem[Krumholz(2006)]{kru06} Krumholz, M., 2006, ApJ, 641, L45.
\bibitem[Kumar et al.(2007)]{kum07} Kumar, M., Davis, C., Grave, J., Ferreira, B. \& Froebrich, D., 2007, MNRAS, 374, 54
\bibitem[Leisawitz et al.(1989)]{lei89} Leizawitz, D., Bash, F. \& Thaddeus, P., 1989, ApJS, 70, 731 
\bibitem[McCaughrean \& Stauffer(1994)]{mcc94} McCaughrean, M., J. \& Stauffer, J., R., 1994, AJ, 108, 1382 
\bibitem[Meyer et al.(1997)]{mey97} Meyer, M., Calvet, N. \& Hillendrand, L., 1997, AJ, 114, 288 
\bibitem[Muench et al.(2000)]{mue00} Muench, A., Lada, E., A. \& Lada, C., J., 2000, ApJ, 533, 358 
\bibitem[Osterloh et al.(1997)]{ost97} Osterloh, M., Henning, Th. \& Launhardt, R., 1997, ApJS, 110, 71 
\bibitem[Ostriker(1999)]{ost99} Ostriker, S., 1999, ApJ, 513, 252 
\bibitem[Plume et al.(1997)]{plu97} Plume, R., Jaffe, D., Evans, N., Martin-Pintado, J. \& Gomez-Gonzalez, J., 1997, ApJ, 476, 730
\bibitem[S\'anchez-Salcedo \& Brandenburg(2001)]{san01} S\'anchez-Salcedo, F. \& Brandenburg, A., 2001, MNRAS, 322, 67
\bibitem[Scalo(1986)]{sca86} Scalo, J. M. 1986, Fundamentals of Cosmic Physics, Vol. 11
\bibitem[Siess et al.(2000)]{sie00} Siess, L., Dufour, E. \& Forestini, M., 2000, A\&A, 358, 593
\bibitem[Spitzer \& Hart(1971)]{spi71} Spitzer, L. \& Hart, M., 1971, ApJ, 164, 399
\bibitem[Stolte et al.(2005)]{sto05} Stolte, A., Brandner, W., Grebel, E., Lenzen, R. \& Lagrange, A-M., 2005, ApJ, 628, 113
\bibitem[Walsh et al.(1998)]{wal98} Walsh, A., Burton, M., Hyland, A. \& Robinson, G.,1998, MNRAS, 301, 640
\end{thebibliography}
\end{document}